\documentclass[aps, prb, a4paper, twocolumn]{revtex4-1}

\usepackage{amsmath}
\usepackage{bm}
\usepackage{epsfig}
\usepackage{amsfonts}
\usepackage{subfigure}
\usepackage{color}
\usepackage{lipsum}
\usepackage{float}
\DeclareMathOperator{\sgn}{sgn}
\DeclareMathAlphabet\mathbfcal{OMS}{cmsy}{b}{n}

\renewcommand{\Re}{{\rm Re}}
\renewcommand{\Im}{{\rm Im}}

\def\e{\begin{equation}}
\def\f{\end{equation}}
\def\_#1{{\bf #1}}

\def\E{\varepsilon}
\def\M{\mu}

\def\.{\cdot}

\begin{document}

\title{Electromagnetic Energy Sink}

\author{C. A. Valagiannopoulos, J. Vehmas, C. R. Simovski, S. A. Tretyakov}
\affiliation{Department of Radio Science and Engineering, School of Electrical Engineering, \\
Aalto University, P.O. Box 13000, FI-00076 Aalto, Finland}

\author{S. I. Maslovski}
\affiliation{
Instituto de Telecomunica\c{c}\~{o}es, Departamento de Engenharia Electrot\'{e}cnica, \\
Universidade de Coimbra P\'{o}lo II, 3030-290 Coimbra, Portugal}

\begin{abstract}
The ideal black body fully absorbs all incident rays, that is, all propagating waves created by arbitrary sources. The known idealized realization of a black body is the perfectly matched layer (PML), widely used in numerical electromagnetics. However, ideal black bodies and PMLs do not interact with evanescent fields existing near any finite-size source, and the energy stored in these fields cannot be harvested. Here we introduce the concept of the ideal conjugate matched layer (CML), which fully absorbs energy of both propagating and evanescent fields of sources acting as an ideal sink for electromagnetic energy. Conjugate matched absorbers have exciting application potentials, as resonant attractors of electromagnetic energy into the absorber volume. We derive the conditions on the constitutive parameters of media which can serve as CML materials, numerically study the performance of planar and cylindrical CML and discuss possible realizations of such materials as metal-dielectric composites.
\end{abstract}

\maketitle


\section{Introduction}
The theoretical concept of the \emph{ideal} black body which completely absorbs all incident rays \cite{GK} is widely used in thermodynamics, optics, and radio engineering. Many important applications would benefit from close-to-ideal absorbers, including light harvesting, thermal emission, stealth, decoupling of antennas and other devices, etc. Motivated by these prospects, significant  research and development efforts have been devoted to realization of absorbers which would offer performance close to that of the ideal black body, see review papers \cite{absorber_review,padilla}. Most of the known results concern planar absorbers, effective for a limited range of incidence angles. Omnidirectional realizations using inhomogeneous lossy media have been proposed \cite{blackhole2,blackhole1} and an experimental test at microwave frequencies has been performed \cite{china_test}. A theoretical possibility to perfectly absorb waves of all polarizations coming from all directions is based on the idea of the \emph{perfectly} matched layer (PML) as a layer which produces no reflections \cite{PML,PML_Gedney,PML_Ziolk,PML_active}. All these concepts imply that the body fully absorbs all propagating waves incident on its surface, emulating the \emph{ideal black body.} But are these objects truly \emph{perfect} absorbers or can we perhaps do even better and extract energy also from evanescent fields created by arbitrary sources? Indeed, the perfectly matched layer produces no reflections and fully absorbs power of all propagating modes (in the ideal case, realizing the properties of the perfect black body), but evanescent fields of sources are not affected by the presence of the perfectly matched layer;  thus, no energy is extracted from them. Basically, within the known paradigm of uniaxial perfectly matched layers \cite{PML_Gedney, PML_active}, there is a conundrum between perfect absorption of all propagating waves and extraction of power from evanescent fields. If the conditions for perfect absorption of propagating fields are satisfied, evanescent modes are not absorbed at all. If the parameters of the absorber are modified so that some power is extracted from evanescent fields of the source, the propagating modes are not any more fully absorbed. In other words, evanescent fields of sources contribute to absorbed power in lossy media, but within all known realization scenarios it appears that if the properties of lossy media are chosen so that all the propagating modes are perfectly absorbed, evanescent fields do not any more interact with the body.

A typical example is a possibility to extract power from near fields of small sources, which has been discussed in the context of hyperbolic media electromagnetics \cite{hyperbolic_review}. Hyperbolic media are artificial materials (metamaterials), whose permittivity tensor is indefinite, so that the real parts of its different diagonal components have the opposite signs. Due to the hyperbolic shape of isofrequency contours, evanescent waves exciting a planar interface of a hyperbolic material sample are partially converted to propagating modes \cite{DipoleLineSource} within the medium and can be effectively absorbed there if the material is lossy \cite{HyperbolicBlackbody, MetamaterialSpontaneousEmission, hyp_shalaev}. However, lossy hyperbolic bodies are not effective absorbers for incident propagating waves. Actually, an interface between free space and a hyperbolic medium can be matched if the anisotropy axis is titled with respect to the interface, but only for one incidence angle \cite{Nefedov}.

In this paper, we study possibilities to create bodies which would act as perfect absorbers for \emph{all} spatial harmonics of incident radiation, both propagating and evanescent. Conceptually, such an ideal absorber is able to extract \emph{infinite} power from a finite-size emitter located at a finite distance, limited only by the power of the source which feeds the emitting object (a generic antenna). This study is motivated by the results of theoretical studies which prove that  a finite-size body, which is \emph{conjugate impedance matched} with respect to its environment \emph{at all spatial spectrum components}, has infinite relative absorptivity when compared to the classical ideal black body \cite{GK} of the same size, e.g. Refs.~\cite{Bobr,unlimited_antenna,SphericalPaper}. However, to the best of our knowledge, only an idealized realization of such a conjugate matched absorber  as a non-uniform, locally isotropic double-negative (DNG) sphere has been proposed in Ref.~\cite{SphericalPaper}. In a sense, such object acts as a perfect sink for all incident spherical waves characterized by arbitrary transverse wave numbers, but its practical realization is challenging because very low levels of material losses are required. Here, we show how one can possibly create absorbers which are nearly perfectly working with both propagating and evanescent incident waves using uniaxial DNG materials, both for planar boundaries and curved surfaces. In this new scenario, there are practical possibilities for realizations as well-studied uniaxial backward-wave structures (see the discussion in Section~\ref{sec:Realization}).

Since our proposed structures are not only able to absorb all propagating waves but can also extract power from the evanescent modes, the total extracted power can go far beyond what can be attained with know metamaterial black holes \cite{blackhole2, blackhole1} and even the ideal black body. In other words, we can go beyond ``the perfect absorber''. From the physical point of view, the conjugate matching condition for evanescent modes is similar to the condition of existence of resonant surface modes, which can help to increase exchange of thermal radiation between two closely located bodies \cite{thermo1, thermo_pendry, thermo1a, thermo2}. However, in the proposed scenario it is possible to realize the optimal conjugate matching for \emph{all} modes, both propagating and evanescent, that is, create the ultimately effective sink for electromagnetic energy from objects located at arbitrary distances. The main focus of this paper is on harvesting energy from small sources in their near-field zone, but we also discuss more general realizations of perfect finite-size absorbers for far-zone sources (plane-wave illumination), that have been also proposed in paper \cite{SphericalPaper}.

\section{Absorption of All Incident Modes}

\subsection{Eliminating Reflection or Maximizing Absorption}
\label{sec:Elimination}

\begin{figure}[ht]
\centering
\subfigure[]{\includegraphics[scale =0.3]{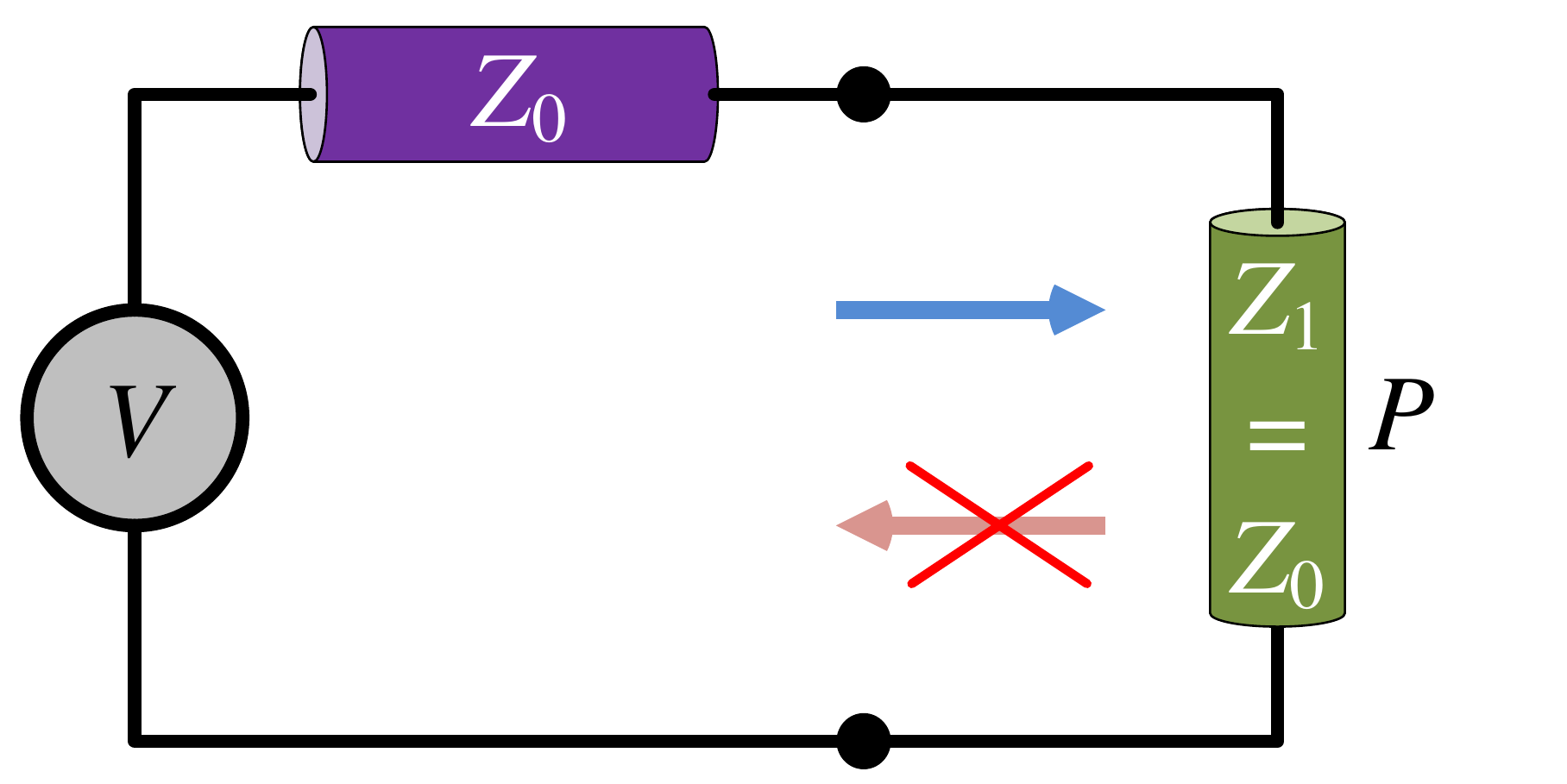}
   \label{fig:Fig1a}}
\subfigure[]{\includegraphics[scale =0.3]{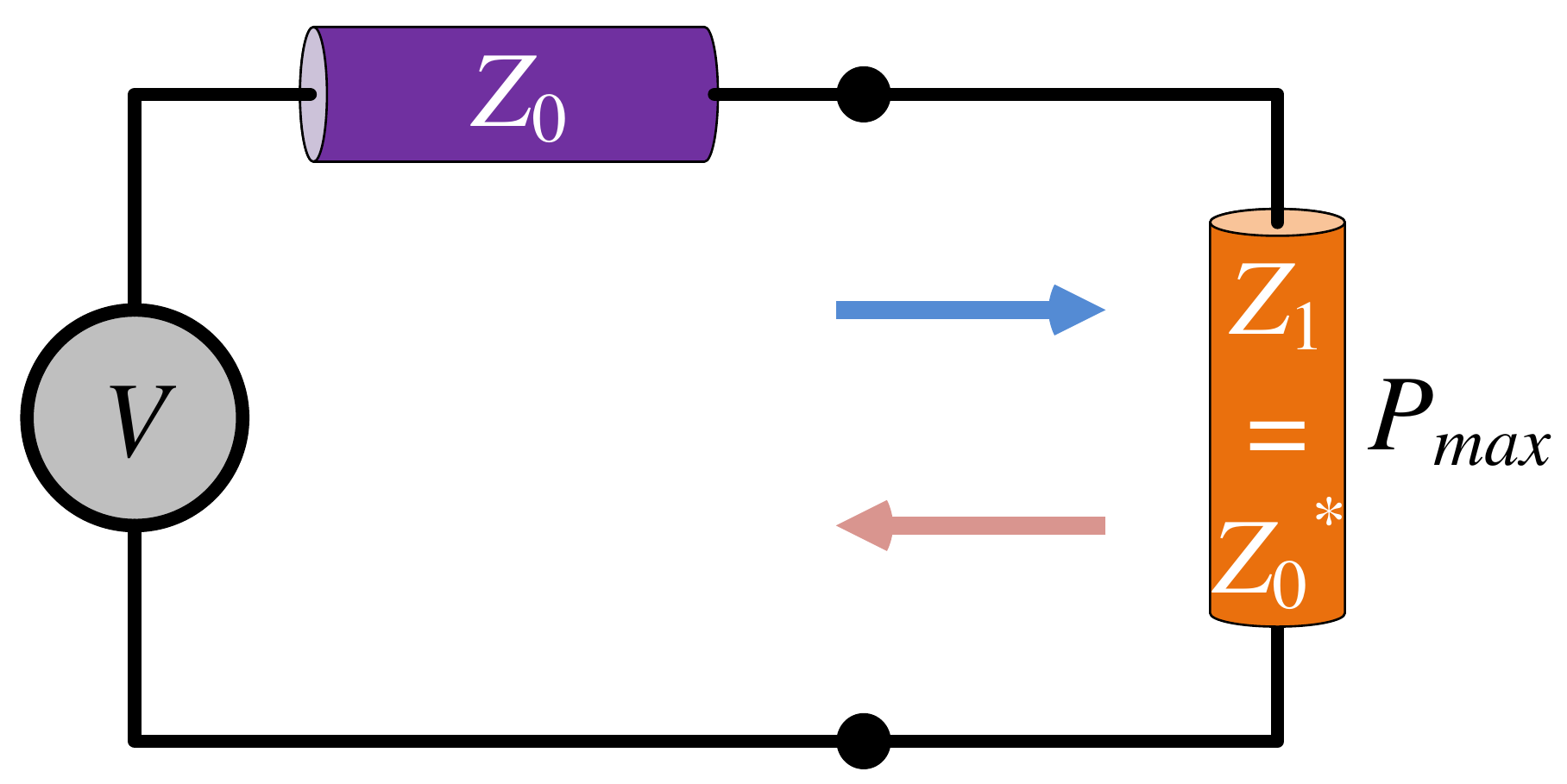}
   \label{fig:Fig1b}}
\caption{Illustration of impedance matching: (a) reflectionless matching where the load impedance $Z_1$ is equal to the source impedance $Z_0$ (the power delivered to the load is $P$), (b) conjugate matching where the load impedance $Z_1$ is equal to the complex conjugate of the source impedance $Z_0^*$ (the power delivered to the load is maximized, $P=P_{\rm max}$).}
\label{fig:Figs1}
\end{figure}

Our starting point is the classical circuit-theory concept of conjugate impedance matching of a load, which is the condition for maximizing the power delivered to the load from a generator. Referring to Fig.~\ref{fig:Figs1}, showing an arbitrary time-harmonic voltage source $V$ with a complex internal impedance $Z_0$, which is loaded by a complex impedance $Z_1$, we can distinguish two different approaches to impedance matching \cite{Pozar}. The first one, illustrated in Fig.~\ref{fig:Fig1a}, minimizes (nullifies) reflections from the load to the source by equalizing the two impedances: $Z_1=Z_0$. In this case the power absorbed in the load equals to (here and in what follows we use rms values for the complex amplitudes of the time-harmonic quantities; the time dependence is $e^{+j\omega t}$, where $j=\sqrt{-1}$):
\begin{equation}
P=\frac{|V|^2}{4|Z_0|^2}\Re[Z_0].
\label{no_refl}
\end{equation}
The second one, illustrated in Fig.~\ref{fig:Fig1b}, maximizes the power
\begin{equation}
P_{\rm max}=\frac{|V|^2}{4\Re[Z_0]}
\label{conj_match}
\end{equation}
that is transferred to the load by satisfying the conjugate matching condition: $Z_1=Z_0^*$. Note that in this case reflections from the load are in general not zero, because the matching condition $Z_1=Z_0$ is satisfied only if these impedances are purely real. Obviously, $P_{\rm max}>P$ if $\Im[Z_0]\ne 0$. With the goal to maximize absorption in the load, the optimal approach is to ensure that the load impedance is conjugate matched with the source impedance.

If the impedances are predominantly reactive (the imaginary parts of $Z_{0,1}$ are large compared with the real parts), the power delivered to the load tends to zero when the load resistance tends to zero [eq. (\ref{no_refl})], while for the conjugate matched load it diverges to infinity [eq.~(\ref{conj_match})]. In the general electromagnetics context, this resonant behavior suggests that conjugate matching wave impedances of two media, in the limit of zero losses, unlimited power can be delivered from sources in one medium into another medium. This fact was noticed in paper \cite{optimization}, in studies of maximizing radiative heat flow between two bodies. As a possible approach to realizing the conjugate match condition for all modes, in paper \cite{optimization} it was proposed to insert dense arrays of thin conducting wires into both bodies (orthogonally to the interface), which makes the wave impedances of a wide range of modes approximately real. This approach is clearly not possible if the sources are in free space, and in this paper we will study means to ensure all-mode conjugate matching by properly selecting the material parameters of the absorbing body.

\begin{figure}[ht]
\centering
\subfigure[]{\includegraphics[scale =0.18]{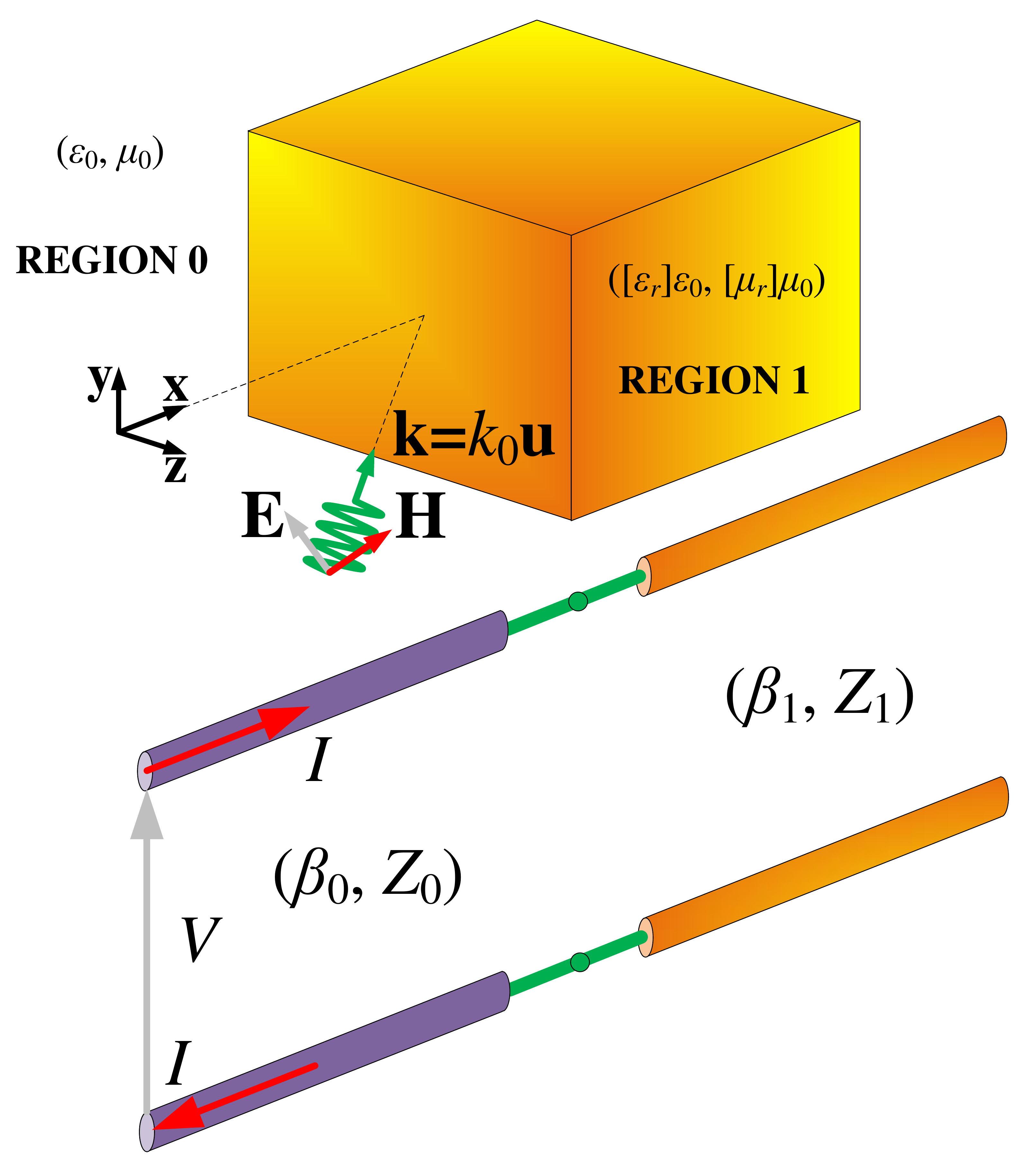}
   \label{fig:Fig2a}}
\subfigure[]{\includegraphics[scale =0.55]{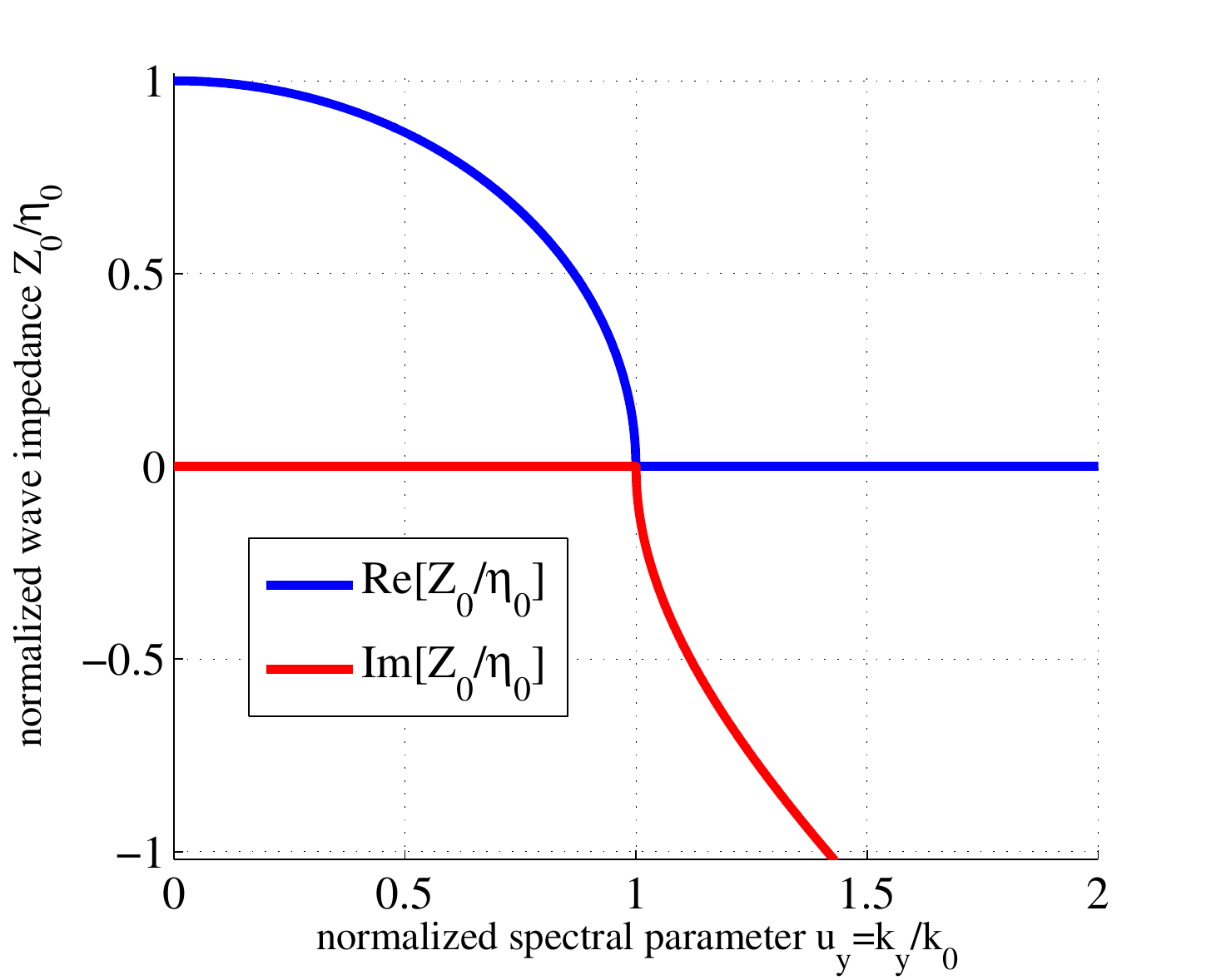}
   \label{fig:Fig2b}}
\caption{(a) A uniaxial material sample (region~1, $\varepsilon_0[\varepsilon_r], \mu_0[\mu_r]$) excited by a plane wave (propagating or evanescent) traveling in free space (region 0, $\varepsilon_0, \mu_0$) along the direction of vector $\textbf{u}$. The corresponding transmission-line model representing the vacuum $(\beta_0, Z_0)$ and the sample $(\beta_1, Z_1)$ is also depicted, together with the utilized Cartesian coordinate system $(x,y,z)$. (b) The variation of the normalized free-space TM impedance $Z_0/\eta_0$ as a function of the normalized spectral parameter $u_y$ (the real and imaginary parts).}
\label{fig:Figs2}
\end{figure}

Let us consider an infinite planar interface between free space and a half-space filled by some lossy material, excited by electromagnetic fields created by some arbitrary sources in free space,  as shown in Fig.~\ref{fig:Fig2a}. We can use the concept of conjugate matching to understand the conditions for maximizing the power absorbed in the medium. To do that, we expand the incident fields into plane-wave spatial Fourier components with the wave vectors $\textbf{k}=k_0\textbf{u}$ ($\textbf{u}=\hat{\textbf{x}}u_x+\hat{\textbf{y}}u_y+\hat{\textbf{z}}u_z)$, where $k_0=2\pi f\sqrt{\varepsilon_0\mu_0}$ is the free-space wavenumber and $\textbf{u}$ is a dimensionless vector. The axis $\hat{\textbf{x}}$ is directed normally to the interface between free space and the absorbing medium and points from free space into the medium. To simplify considerations, without loss of generality we will consider fields of only one polarization --- the TM polarization (magnetic field possesses a sole $\hat{\textbf{z}}$ component and electric field vector lies on the $xy$ plane) --- and assume that the fields do not depend on the coordinate $z$. The normal to the interface component of the propagation constant in free space reads $\beta_0=-jk_0\sqrt{u_y^2-1}=-jk_0s_0(u_y)$, where $u_y\in\mathbb{R}$. When $|u_y|<1$, the excitation corresponds to propagating waves and when $|u_y|>1$, we deal with exponentially decaying evanescent modes of free space. The corresponding TM wave impedance in vacuum (the ratio of the tangential to the interface field components) is expressed as
\begin{equation}
Z_0=-j\eta_0s_0(u_y)=-j\eta_0\sqrt{u_y^2-1},
\label{Z0}
\end{equation}
where $\eta_0=\sqrt{\mu_0/\varepsilon_0}$ is the free-space wave impedance. The branches of all square roots are defined as having positive real parts.

In view of our goal to conjugate match the impedances of \emph{all} modes, we demand the uniaxial symmetry of the material tensors of the absorbing medium, with the axis directed normally to the interface (along $\hat{\textbf{z}}$), to ensure a possibility of matching to the isotropic free space. Let us write the relative permittivity and permeability tensors as
\begin{eqnarray}
[\varepsilon_r]=\left[\begin{array}{ccc}
\varepsilon_{rn} & 0             & 0 \\
0             & \varepsilon_{rt} & 0 \\
0             & 0             & \varepsilon_{rt}
\end{array}\right]
\label{eq:PermittivityTensor}~,~
[\mu_r]=\left[\begin{array}{ccc}
\mu_{rn} & 0             & 0 \\
0             & \mu_{rt} & 0 \\
0             & 0             & \mu_{rt}
\end{array}\right].
\label{eq:PermeabilityTensor}
\end{eqnarray}
The propagation constants of transverse magnetic (TM-polarized) plane waves in the medium read \cite{modeboo}: $\beta_1=-jk_0s(u_y)$ with
\begin{eqnarray}
s(u_y)=\sqrt{\frac{\varepsilon_{rt}}{\varepsilon_{rn}}u_y^2-\varepsilon_{rt}\mu_{rt}},
\label{eq:SQuantity}
\end{eqnarray}
where the real part of the square root is positive, ensuring decaying waves along the positive direction of the axis $\hat{\textbf{z}}$ (away from the sources). The corresponding impedance equals
\begin{equation}
Z_1=-j\eta_0\frac{s(u_y)}{\varepsilon_{rt}}.
\label{Z1}
\end{equation}

To understand the conditions for the optimal absorption of all modes we can use the concept of  conjugate impedance matching for all plane-wave components of the fields. In the configuration of Fig.~\ref{fig:Figs2}, the load impedance is the wave impedance of a certain mode in the medium and the source impedance $Z_0$ is the impedance of the corresponding (the same $u_y$) mode in free space. The transmission-line model with voltage $V$ (corresponding to the electric field $\textbf{E}$) and current $I$ (measuring the magnetic field $\textbf{H}$) is also depicted at the bottom of Fig.~\ref{fig:Fig2a}, where the two lines with different propagation constants $(\beta_0, \beta_1)$ and wave impedances $(Z_0, Z_1)$ represent the source and the load, respectively.

The variation of the normalized free-space wave impedance (with infinitesimally small losses added to ensure the correct choice of the square-root branch) of TM waves $Z_0/\eta_0$ as a function of the relative tangential component of the propagation constant $u_y=(\textbf{k}\cdot \hat{\textbf{y}})/k_0=k_y/k_0$ is shown in Fig.~\ref{fig:Fig2b} (only positive $u_y$ are considered since all the impedance functions are even with respect to $u_y$). It is apparent that the quantity $Z_0$ is positive real for $|u_y|<1$ (propagating modes) and negative imaginary for $|u_y|>1$ (evanescent modes). With the objective to maximize absorption of electromagnetic energy by the medium, we will next look for such constituent parameters $(\varepsilon_{rt}, \mu_{rt}, \varepsilon_{rn})$ that $Z_1=Z_0^*$ for all real values of $u_y$.

\subsection{Double-Negative Conjugate Matched Layer}

Given the fact that the TM wave impedance of vacuum (\ref{Z0}) is real for $|u_y|<1$, the reflectionless matching is identical to conjugate matching for the propagating waves. Therefore, for that part of the incidence spectrum, we are searching for a uniaxial medium that does not reflect incoming traveling fields. Such a medium is the famous uniaxial perfectly matched layer (PML) with \cite{PML_Gedney}
\begin{equation}
\varepsilon_{rt}=\mu_{rt}=\frac{1}{\varepsilon_{rn}}.
\label{PML_rule}
\end{equation}
Indeed, substitution of (\ref{PML_rule}) into (\ref{Z1}) and comparison with (\ref{Z0}) shows that $Z_1=Z_0$ for any value of $u_y$. However, the fact that a half space with these characteristics exhibits no reflections also for evanescent modes, namely that $Z_1=Z_0$ also for $|u_y|>1$, does not fit well with the maximal absorption goal.

\begin{figure}[ht]
\centering
\subfigure[]{\includegraphics[scale =0.4]{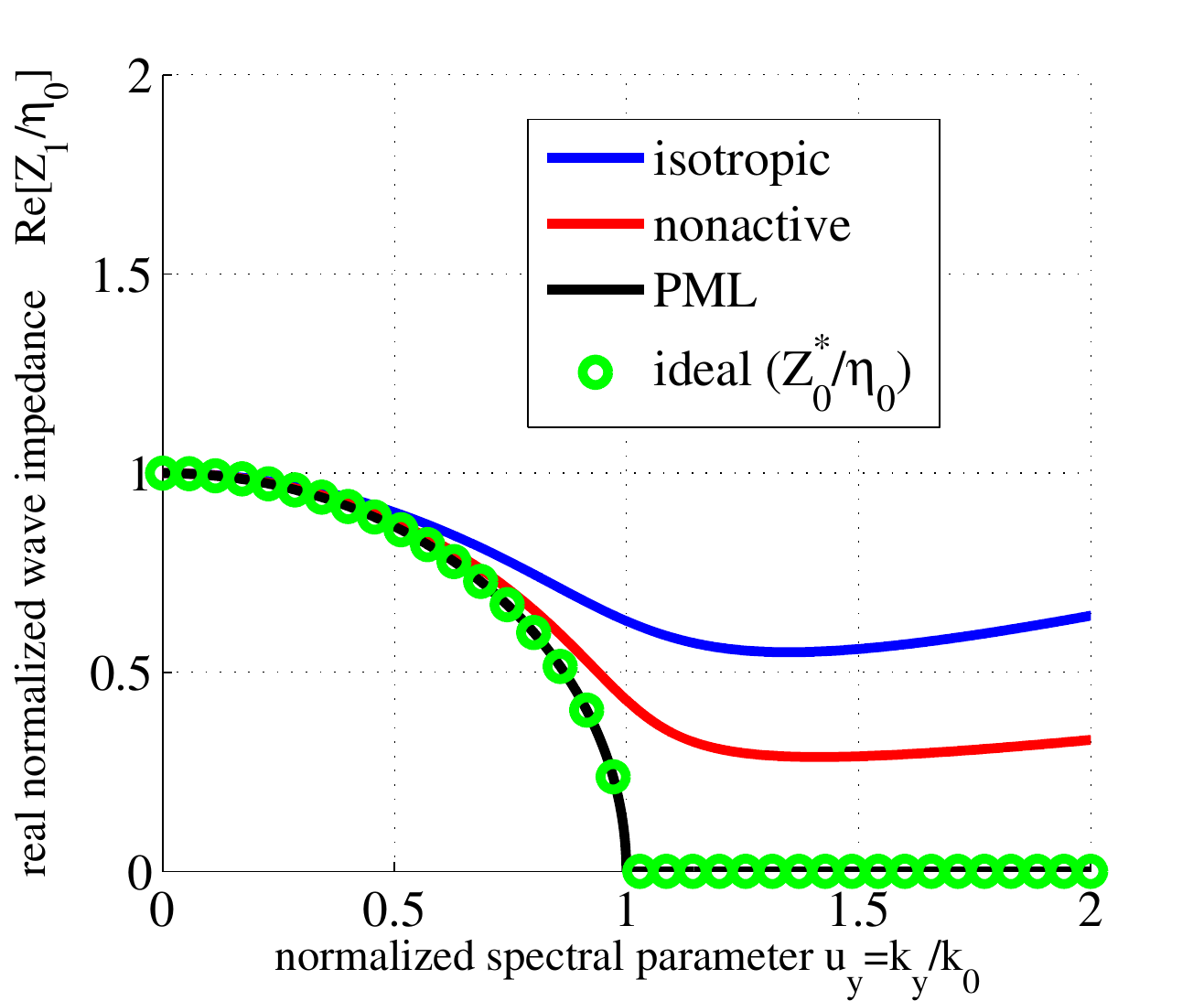}
   \label{fig:Fig3a}}
\subfigure[]{\includegraphics[scale =0.4]{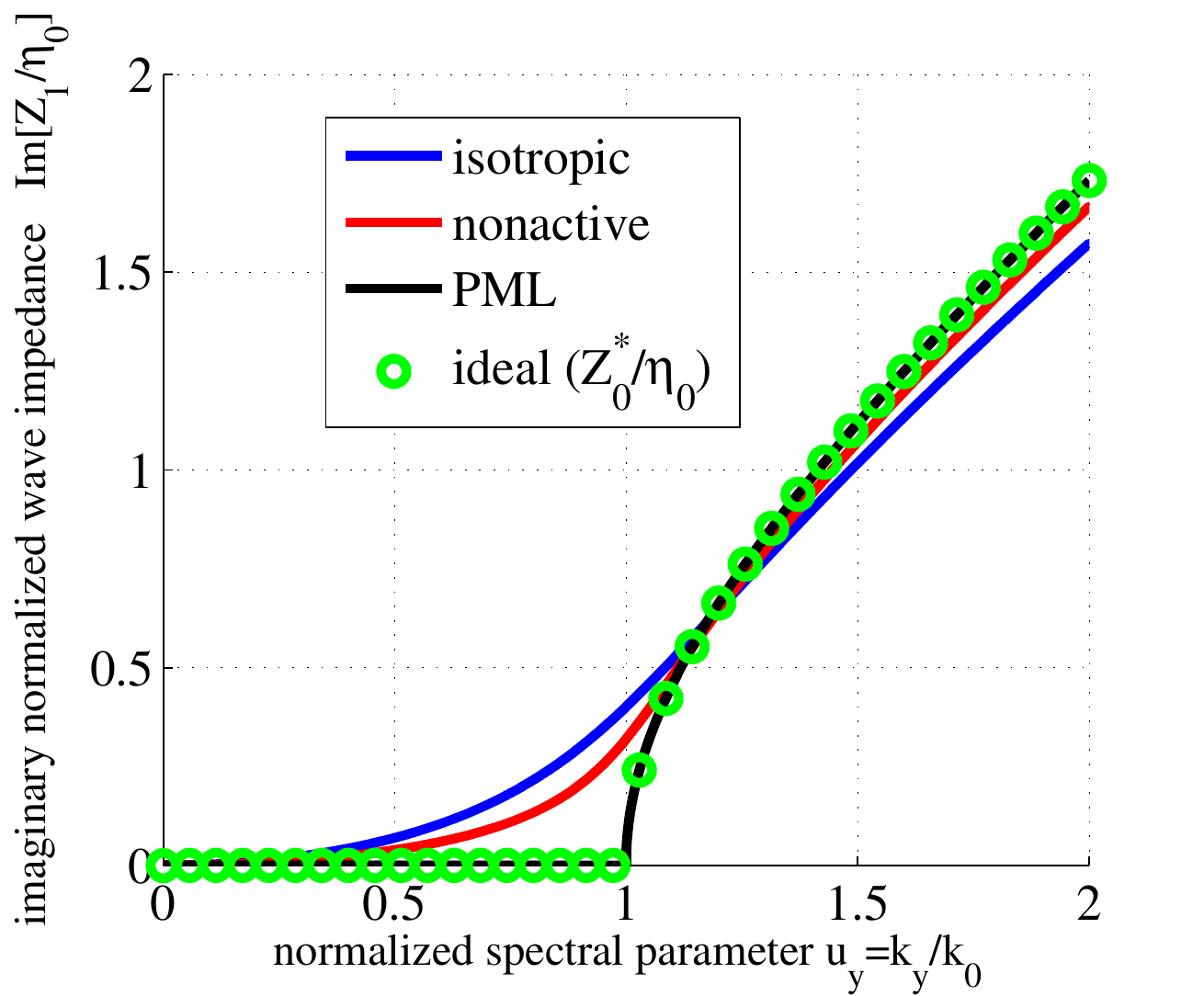}
   \label{fig:Fig3b}}
\caption{The variation of: (a) the real parts and (b) the imaginary parts of the normalized TM impedance $Z_1/\eta_0$ for a uniaxial medium as a function of the normalized spectral parameter $u_y$. We examine three cases (routes) based on the PML concept: (i) normally matched double-negative (DNG) isotropic medium with $\varepsilon_{rn}=\varepsilon_{rt}=\mu_{rn}$ (isotropic route), (ii) DNG lossy PML-type medium with a lossless normal component, where: $\varepsilon_{rn}=\Re\left[1/\varepsilon_{rt}\right]=\Re\left[1/\mu_{rt}\right]$, so that none of the three constituent parameters $(\varepsilon_{rt}, \mu_{rt}, \varepsilon_{rn})$ is active (nonactive route) and (iii) DNG lossy PML-type medium with $\varepsilon_{rn}=1/\varepsilon_{rt}=1/\mu_{rt}$ (active $\varepsilon_{rn}$, PML route). In all the cases $\varepsilon_{rt}=\mu_{rn}=-1-j0.3$.}
\label{fig:Figs3}
\end{figure}

It is known that a planar interface with an isotropic Veselago medium ($\varepsilon_r=\mu_r=-1$) is conjugate matched with free space for all modes \cite{modeboo}, which is the enabling property for the perfect lens operation \cite{DNG}. However, this material is lossless for both propagating and evanescent waves. If we introduce small losses in order to extract energy from evanescent waves, the perfect matching for propagating modes is destroyed. The power generated by evanescent fields can be increased by making losses small [eq.~(\ref{conj_match})], but at the same time absorption of propagating modes gets smaller and smaller (the imaginary part of the propagation constant tends to zero). To overcome this deficiency, we propose to use double-negative materials whose parameters satisfy (\ref{PML_rule}), but $\Re[\varepsilon_{rt}]=\Re[\mu_{rt}]<0$. Obviously, the real part of the normal component of the permittivity is also negative: $\Re[{\varepsilon_{rn}}]=\Re\left[1/\varepsilon_{rt}\right]<0$.
Using such materials, conjugate matching for all propagating modes can be realized also in lossy configurations, where all propagating plane waves quickly decay inside the absorbing medium. This property allows us to expect that we can effectively extract power from \emph{both} propagating and evanescent waves. We will call such structures (double-negative) \emph{conjugate matched layers} (CML).

It is needless to say that if the constituent parameters are purely real, no absorption takes place. Accordingly, we should add some losses to the transversal components; in this situation, all propagating waves exponentially decay and give up their energy to the medium. However, in this case the normal permittivity components $\varepsilon_{rn}=1/\varepsilon_{rt}=1/\mu_{rt}$ are inevitably active. Indeed, the sign of the imaginary parts of $\varepsilon_{rn}$ and $\mu_{rn}$ is the opposite to that of the imaginary parts of $\varepsilon_{rt}$ and $\mu_{rt}$, as the material parameters comply with the PML conditions (\ref{PML_rule}). Thus, the material parameters of the ideal double-negative conjugate matched layer do not have any active components only in limiting cases when losses tend to zero or the real parts of the transverse components are very large.

Let us next show that double-negative materials whose parameters satisfy (\ref{PML_rule}) are indeed conjugate matched to free space for all modes. For propagating waves ($|u_y|<1$) we have  $s(u_y)=\sqrt{\varepsilon_{rt}^2(u_y^2-1)}=\varepsilon_{rt}\sqrt{u_y^2-1}$, which corresponds to the normal component of the propagation constant $\beta_1=-jk_0s(u_y)$ having a negative real part (propagating backward waves, as in any double-negative material). The chosen branch $\Re(\sqrt{\cdot}\,)>0$ corresponds to the proper direction of the power flow from the sources into the absorbing medium. The corresponding impedance reads $Z_1=-j\eta_0s(u_y)/\varepsilon_{rt}=-j\eta_0\sqrt{u_y^2-1}= Z_0=Z_0^*$, because this is a real number. For evanescent modes ($|u_y|>1$) we get $s(u_y)=\sqrt{\varepsilon_{rt}^2(u_y^2-1)}=-\varepsilon_{rt}\sqrt{u_y^2-1}$, where the sign is chosen so that the fields decay away from the sources in free space. The corresponding impedance becomes $Z_1=-j\eta_0s(u_y)/\varepsilon_{rt}=j\eta_0\sqrt{u_y^2-1}= Z_0^*$.

In Figs.~\ref{fig:Figs3} we show the variation of the real (Fig.~\ref{fig:Fig3a}) and imaginary (Fig.~\ref{fig:Fig3b}) parts  of the normalized wave impedance $Z_1/\eta_0$ (\ref{Z1}) as functions of $u_y$ for materials with the parameters approximating the ideal values of the CML layer. We examine three possible routes towards CML: (i) An isotropic DNG material with $\varepsilon_{rn}=\varepsilon_{rt}=\mu_{rn}$, which has the CML properties in the limit $\varepsilon_{rn}=\varepsilon_{rt}=\mu_{rn}\rightarrow -1$; this approach is labeled as the ``isotropic route''. (ii) A lossy uniaxial double-negative material with  the lossless normal component of the permittivity, satisfying  $\varepsilon_{rn}=\Re\left[1/\varepsilon_{rt}\right]=\Re\left[1/\mu_{rt}\right]$, which formally has the ideal CML properties in the limit $\Re[\varepsilon_{rt}]=\Re[\mu_{rt}]\rightarrow -\infty$; this approach is labeled as the ``nonactive route''. Finally, (iii) a double-negative lossy CML medium with $\varepsilon_{rn}\cong 1/\varepsilon_{rt}=1/\mu_{rt}$; this approach is labeled as the ``PML route''. The impedances $Z_1$ are compared with the ideal complex conjugate wave impedance $Z_0^*$ of free space: The equality of the two quantities  guarantees maximal absorption.

It is clear that the choice of negative real parts for $(\varepsilon_{rt}, \mu_{rt}, \varepsilon_{rn})$ is successful in terms of ensuring the desired sign of $\Im[Z_1]$; one should take into account that for the usual PML with positive real constituent parameters (double-positive, DPS), we obtain $Z_1=Z_0$ for all real $u_y$ and thus $\Im[Z_1]<0$, as seen in Fig.~\ref{fig:Fig2b}. It is also apparent that for both real and imaginary parts, the uniaxial material with lossless $\varepsilon_{rn}$ performs better than the isotropic material. Most importantly, the double-negative lossy CML corresponds to \emph{exact conjugate matching}, the fact that verifies our initial assumption. Note that such a selection of a ``PML'' with $\Re[\varepsilon_{rt}]<0$ and $\Im[\varepsilon_{rt}]<0$ changes (compared to the normal DPS PML) only the wave impedance of evanescent modes since $Z_1$ is purely real along the propagating spectrum.

One can say that we \emph{misapply} the concept of PML since the choice of double-negative components inflicts strong reflections $(Z_1=Z_0^*)$ for the evanescent modes, namely for $|u_y|>1$, contrary to the usual reflectionless matching property of DPS PML $(Z_1=Z_0)$. However, within our paradigm, all propagating waves are fully absorbed, and evanescent fields, in the limit of ideal conjugate matching, generate infinitely strong fields in the medium and deliver infinite power to the absorber in the limit of $\Re[Z_1]\rightarrow 0$ [Eq.~(\ref{conj_match})], in the assumption that the illuminating antennas  are fed by ideal voltage or current sources capable to supply unlimited power.

\begin{figure}[ht]
\centering
{\includegraphics[scale =0.39]{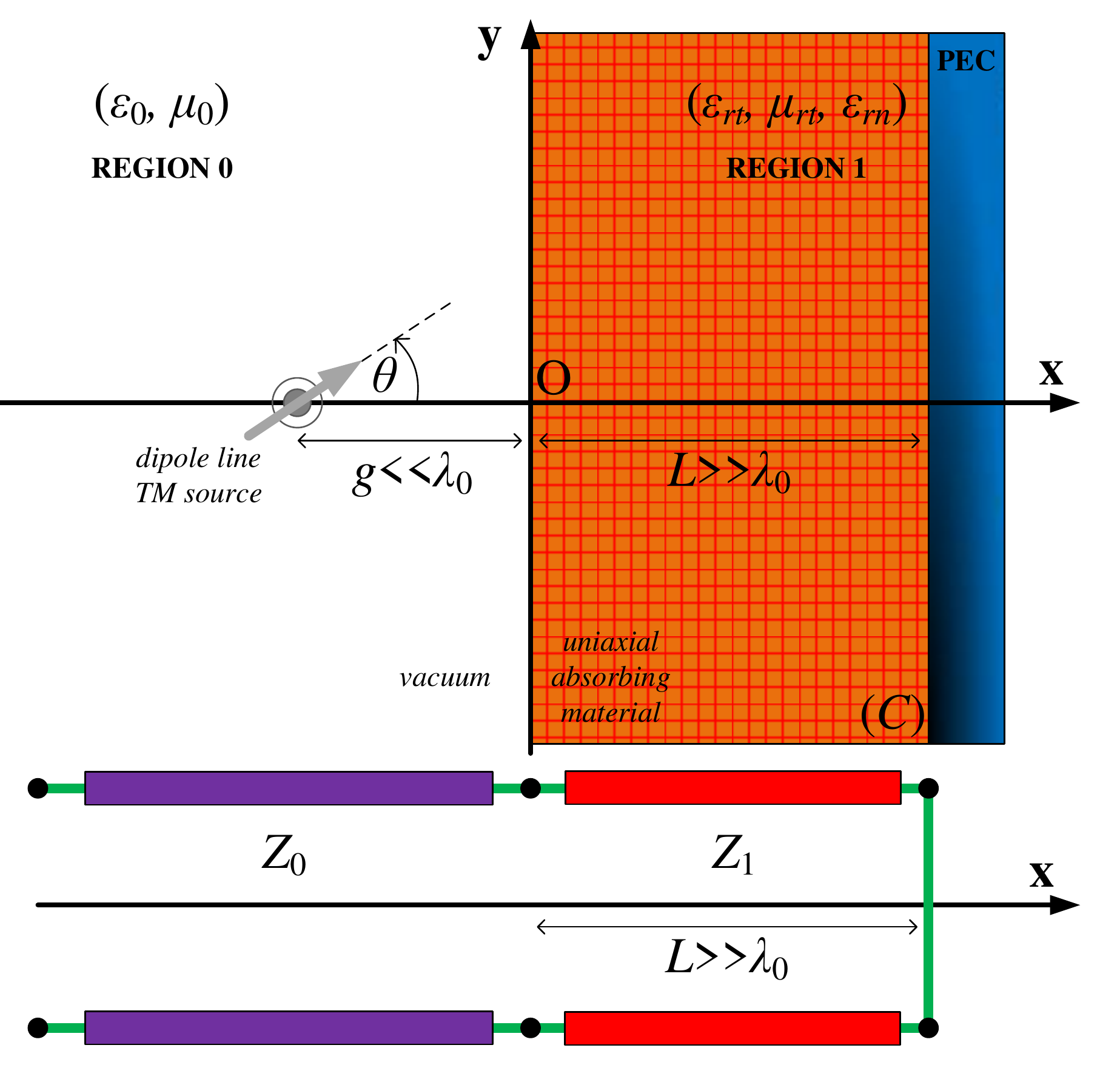}
\caption{A grounded uniaxial slab of thickness $L$ is excited by an electric dipole line source forming an angle $\theta$ with the horizontal axis, placed at a close distance $g$ from the boundary. The corresponding transmission-line model is also depicted.}
\label{fig:Fig4a}}
\end{figure}

\section{Infinite Planar Configuration}
\label{sec:Planar}

\subsection{Incident Field and Absorbed Power}
To test and demonstrate the absorbing efficiency of double-negative
conjugate matched layers, we consider the planar structure depicted in
Fig.~\ref{fig:Fig4a}. A grounded slab of thickness $L$ (region 1)
filled with a uniaxial medium $(\varepsilon_{rt}, \mu_{rt},
\varepsilon_{rn})$ is excited by an electric dipole line source
\cite{DipoleLineSource} located at the point $(x,y)=(-g,0)$ in the
vacuum region (region 0). The dipole moment per unit length of the
line is denoted as $\_p_l$. The dipoles form an angle $\theta$ with
the horizontal axis $\hat{\textbf{x}}$. The distance between the slab
and the source is much smaller than the free-space wavelength (in
order for the evanescent waves not to vanish before reaching the
sample), and is denoted by $g$. To study absorption of evanescent
fields, we choose such a source since its evanescent spectral
components are much more pronounced compared to that of more typical
cylindrical wave sources such as lines of electric or magnetic
currents.~\cite{Evanes} The problem is effectively two-dimensional
(2D), due to $z$-independence, and the magnetic field has a single
$\hat{\textbf{z}}$ component.

We write the solutions in spectral domain, as a plane-wave
decomposition in the form $\exp(-jk_y y)$. The expression
(\ref{eq:TransformedMagneticField}) of the Fourier-transformed
incident magnetic field $\mathbfcal{H}_{\rm 0,inc}(x,
k_y)=\hat{\textbf{z}}\mathcal{H}_{\rm 0,inc}(x, k_y)$ obtained in
Appendix can be evaluated at $x=0$ as follows:
\begin{equation}
\mathcal{H}_{\rm 0,inc}(x=0, k_y)=\frac{j\omega p_l}{2}e^{-j\beta_0g}\left[\frac{k_y}{\beta_0}\cos\theta-\sin\theta\right],
\label{ExcitationField}
\end{equation}
where $p_l=|\textbf{p}_l|$. This field excites the uniaxial slab
located at $x>0$. We note that identical plane-wave field components
are generated by sheets of surface electric current $\textbf{J}_s=
-2\hat{\textbf{y}}\mathcal{H}_{\rm 0,inc}(0, k_y)$ placed right in
front of the grounded slab. Therefore, the excitation of the slab by
the line of electric dipoles (Fig.~\ref{fig:Fig4a}, top) can be
equivalently represented by the transmission-line model shown at the
bottom of Fig.~\ref{fig:Fig4a}, with the external current source
$\textbf{J}_s$ inserted at $x=0$. Using this circuit model, the power
delivered to the slab by each incident spatial harmonic is
straightforwardly calculated as:
\begin{equation}
p(k_y)=\frac{4|\mathcal{H}_{\rm 0,inc}(0, k_y)|^2 |Z_0(k_y)|^2}{|Z_0(k_y)+Z_{\rm in}(k_y)|^2}\Re[Z_{\rm in}(k_y)].
\label{PowerSpectralDensity}
\end{equation}
As above, by $Z_0$ we denote the free-space wave impedance, and $Z_{\rm in}$ is the input impedance of the metal-backed uniaxial slab evaluated from the corresponding transmission-line model as $ Z_{\rm in} =jZ_1\tan(\beta_1 L).$ Finally, the total power absorbed in the slab is found by integration over all the wavenumber spectrum:
\begin{widetext}
\begin{equation}
P=\frac{1}{2\pi}\int_{-\infty}^{+\infty}p(k_y)dk_y=\frac{\omega^2p_l^2}{2\pi}\int_{-\infty}^{+\infty}e^{-2\beta''_0(k_y)g}\left|\frac{k_y}{\beta_0(k_y)}\cos\theta-\sin\theta\right|^2
\frac{\Re[Z_{\rm in}(k_y)]}{\left|1+\frac{Z_{\rm in}(k_y)}{Z_0(k_y)}\right|^2}dk_y,
\label{AbsorbedPower}
\end{equation}
\end{widetext}
Here it is denoted $\beta_0'' = -\mathop{\rm Im}(\beta_0)$ (for propagating waves, $\beta_0''= 0$, and for evanescent waves, $\beta_0''=\sqrt{k_y^2-k_0^2}$).

\subsection{Important Limiting Cases}
Let us now consider some limiting cases under the simplifying
assumption of a half space emulating an electrically thick absorbing
slab (valid when $\beta_1''L\gg 1$, where $\beta_1''=-\Im[\beta_1]$). In this
case, the input impedance of the metal-backed uniaxial slab approaches
its wave impedance: $Z_{\rm in}\cong Z_1$. In our study, the material
parameters of the uniaxial layer are chosen so that the conjugate
matching condition is closely approached: $Z_1\cong Z_0^*$. When
$|k_y|<k_0$ (propagating waves), $Z_0=\eta_0\beta_0/k_0$ is purely
real and, thus, $Z_1 \cong Z_0$. Therefore, the power delivered by the
propagating modes to the conjugate-matched layer can be calculated as:
\begin{equation}
P_0\cong \frac{\eta_0\omega^2p_l^2}{8\pi}\int_{-k_0}^{k_0}\left|\frac{k_y}{\beta_0}\cos\theta-\sin\theta\right|^2\frac{\beta_0}{k_0}dk_y=\frac{\M_0\omega^3p_l^2}{16}.
\label{PropagatingAbsorption}
\end{equation}
Note that this result is independent of the angle $\theta$ because
(\ref{PropagatingAbsorption}) is nothing more than $1/2$ of the total
power emitted by a line of dipoles in unbounded free space. This power also
equals the total amount of power absorbed by an ideal PML layer (or,
equivalently, by a perfect black body) in the structure shown in
Fig.~\ref{fig:Fig4a}.

When $|k_y|>k_0$ (evanescent waves), $Z_0$ is purely imaginary and,
thus, fulfilling the conjugate matching condition results also in
almost purely imaginary $Z_1$: $Z_1\cong Z_{\rm in}\cong
-j\Im[Z_0]$. In this case, the denominator of (\ref{AbsorbedPower})
approaches zero. Therefore, in order to analyze this limiting case we
must consider a particular model for the material parameters of the
uniaxial layer. Here we select the PML route with $\E_{rt} = \M_{rt} =
a - jb$ and $\E_{rn} = 1/\E_{rt} - j\delta$, where real $\delta$ is
such that $|\delta|\ll 1$. Performing the Taylor expansion with
respect to small $\delta$ under the aforementioned assumption
$|k_y|>k_0$ and keeping only the first-order term, we obtain the
following approximation for $Z_1$:
\begin{equation}
Z_1(k_y)=\frac{\eta_0\beta_1}{\E_{rt}k_0}\cong\sgn(a)Z_0(k_y)\left[1+j\delta~\frac{(a-jb)k_y^2}{2\left(k_y^2-k_0^2\right)}\right].
\label{eq:Z1Approximation}
\end{equation}
Separating the real and imaginary parts and taking into account that
$\sqrt{(a-jb)^2}=\sgn(a)(a-jb)$, we come to the following formula:
\begin{widetext}
\begin{equation}
\frac{\Re[Z_1(k_y)]}{\left|1+Z_1(k_y)/Z_0(k_y)\right|^2}\cong\eta_0\frac{|a|k_y^2}{2k_0\sqrt{k_y^2-k_0^2}}
\frac{\delta}{\left[\frac{a\delta k_y^2}{2\left(k_y^2-k_0^2\right)}\right]^2+\left[1+\sgn(a)+b\sgn(a)\frac{\delta k_y^2}{2\left(k_y^2-k_0^2\right)}\right]^2}~,\quad \delta\rightarrow 0.
\label{FractionApproximation}
\end{equation}

Now, we can substitute (\ref{FractionApproximation}) into
(\ref{AbsorbedPower}) and calculate the integral over the
evanescent part of the spectrum ($|k_y|>k_0$).  This calculation
results in the following approximate expression for the power absorbed from the evanescent modes:
\begin{equation}
P_{\rm evan}\cong \frac{8|a|}{k_0^2\pi}P_0 \int_{k_0}^{+\infty}\frac{k_y^2\left(k_y^2-k_0^2\sin^2\theta\right)}{\left(k_y^2-k_0^2\right)^{3/2}}e^{-2g\sqrt{k_y^2-k_0^2}}
\frac{\delta}{\left[1+\sgn(a)\right]^2+\delta^2\left[\frac{k_y^2 |\varepsilon_{rt}|}{2\left(k_y^2-k_0^2\right)}\right]^2}dk_y~,\quad \delta\rightarrow 0.
\label{EvanescentAbsorbedPower}
\end{equation}
\end{widetext}

In the denominator of the integrand of (\ref{EvanescentAbsorbedPower})
we have dropped the term proportional to $\delta$, because in the DNG
case $(a<0)$ it is multiplied by $1+\sgn(a)=0$ and in this case the
first nonvanishing term is the second-order term in $\delta$ [this
term is present in (\ref{EvanescentAbsorbedPower})], while in the DPS
case $(a>0)$ the main term is $\left[1+\sgn(a)\right]^2=4 \gg \delta
\gg \delta^2$. Equation (\ref{EvanescentAbsorbedPower}) demonstrates
the dramatic influence of the sign of the real part of
$\varepsilon_{rt}=\mu_{rt}$. In particular, for the CML case when $a$
is negative, the denominator of (\ref{EvanescentAbsorbedPower}) is
approximately proportional to $\delta^2$, and the following formula
results from (\ref{EvanescentAbsorbedPower}):
\begin{equation}
\frac{P_{\rm evan}}{P_0}\cong\frac{32|a|}{\pi\delta |\varepsilon_{rt}|^2}\int_{0}^{+\infty}e^{-2k_0gs_0}\frac{s_0^2\left(s_0^2+\cos^2\theta\right)}{\left(1+s_0^2\right)^{3/2}}ds_0,
\label{eq:PDNGevanescent}
\end{equation}
where a change of the integration variable $s_0=\sqrt{k_y^2-k_0^2}/k_0=\sqrt{u_y^2-1}$ has been made. We notice that the absorbed power is proportional to $1/\delta$, which means that for small $\delta$ we have huge magnitudes either of absorbed $(\delta>0)$ or emitted $(\delta<0)$ powers. Such a limiting regime is explained by the singularity of (\ref{conj_match}) for $\Re[Z_1]\rightarrow 0$.

Expression (\ref{EvanescentAbsorbedPower})
can be  applied also for the nonactive
route when $\E_{rn} = \mathop{\rm Re}(1/\E_{rt})$. Indeed, when
$\delta = b/(a^2 + b^2) \ll 1$, the imaginary part of $\E_{rn}$
vanishes and the normal component of the permittivity dyadic acquires
the value $\E_{rn} = a/(a^2+b^2) = \mathop{\rm
  Re}(1/\E_{rt})$. Respectively, Eq.~(\ref{EvanescentAbsorbedPower})  becomes
\begin{equation}
\frac{P_{\rm evan}}{P_0}\cong\frac{32|a|}{\pi b}\int_{0}^{+\infty}e^{-2k_0gs_0}\frac{s_0^2\left(s_0^2+\cos^2\theta\right)}{\left(1+s_0^2\right)^{3/2}}ds_0.
\label{eq:PNAGevanescent}
\end{equation}
In this case, the absorbed power
increases with the decrease of $b$ and the increase of $|a|$.

On the other hand, for the conventional DPS PML case, expression (\ref{EvanescentAbsorbedPower}) takes the following form:
\begin{equation}
\frac{P_{\rm evan}}{P_0}\cong\frac{2a}{\pi}\delta\int_{0}^{+\infty}e^{-2k_0gs_0}\frac{\sqrt{s_0^2+1}\left(s_0^2+\cos^2\theta\right)}{s_0^2}ds_0.
\label{eq:PDPSGevanescentDivergent}
\end{equation}
In this case, when $\delta\rightarrow 0$ (approaching the ideal impedance matching condition), the relative power delivered to the slab by the evanescent waves tends to zero. This result confirms that within the know PML scenario, the goal of full absorption of propagating waves is incompatible with the goal of extraction power from evanescent fields.

Note that the above approximate formula makes sense only when $\theta
= \pi/2$, because only in this case the integral
(\ref{eq:PDPSGevanescentDivergent}) converges. The divergence of the
integral (\ref{eq:PDPSGevanescentDivergent}) at any other values of
$\theta$ is an artifact of the Taylor expansion approximation
(\ref{eq:Z1Approximation}). This approximation is not applicable when
$k_y$ approaches $k_0$ (i.e., when $s_0 \rightarrow 0$). However,
direct evaluation of the integrand in (\ref{EvanescentAbsorbedPower})
shows no singularity at $k_y = k_0$, even when $\theta \neq
\pi/2$. Therefore, when $\theta \neq \pi/2$, the lower limit of the
integration in Eq.~(\ref{eq:PDPSGevanescentDivergent}) must be
replaced by $\sigma \sim \sqrt{\delta|\E_{rt}|/2}$, because at this
point the first term of the Taylor expansion
(\ref{eq:Z1Approximation}) is about unity:
$\frac{|\varepsilon_{rt}|k_y^2}{2(k_y^2-k_0^2)}\delta\cong 1$. In this
case, Eq.~(\ref{eq:PDPSGevanescentDivergent}) qualitatively agrees
with the accurate result~(\ref{AbsorbedPower}) (when only the
contribution of the domain $|k_y| > k_0$ is considered), with the
dominant contribution to the integral coming from a narrow region in
the vicinity of the lower limit. Therefore, when $\theta \neq \pi/2$
and $k_0g \ll 1$,
\begin{equation}
\frac{P_{\rm evan}}{P_0}\cong\frac{2a}{\pi}\delta\int_{\sigma}^{+\infty}\frac{\cos^2\theta}{s_0^2}ds_0=\sqrt{\frac{8\delta}{|\varepsilon_{rt}|}}\frac{a\cos^2\theta}{\pi},
\label{eq:PDPSGevanescentConvergent}
\end{equation}
i.e., the
power delivered by the evanescent modes in this case is roughly
proportional to $\sqrt{\delta}$ and also vanishes when
$\delta\rightarrow 0$.

\subsection{Performance of the Proposed Structure}
\label{sec:performance}

Here we characterize the proposed conjugate matched layers qualitatively, studying the ratio of the total absorbed power and the power which can be harvested only from the propagating part of the spatial spectrum of the excitation in the layer of the infinite thickness (\ref{PropagatingAbsorption}). Basically, this normalized parameter is the ratio of the power absorbed by our structures and the power absorbed in the \emph{classical ideal black body} at the same position and excited by the same source. In particular, the output quantity is the logarithm (base 10) of this ratio $\log\left(P/P_0\right)$ since we are expecting huge magnitude variations of $|P|$, as discussed in the previous section.

\begin{figure}[ht]
\centering
\subfigure[]{\includegraphics[scale =0.28]{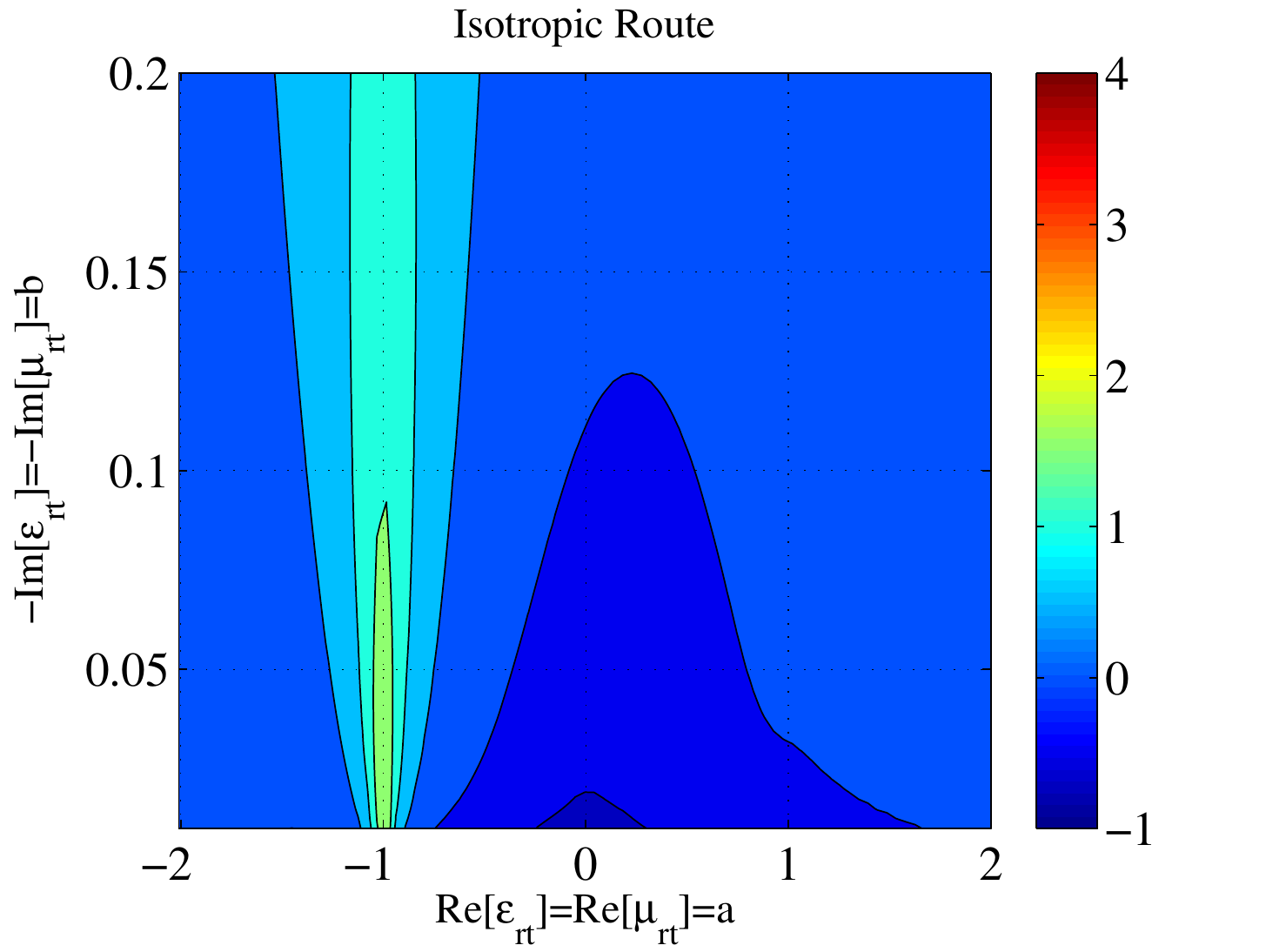}
   \label{fig:Fig5a}}
\subfigure[]{\includegraphics[scale =0.28]{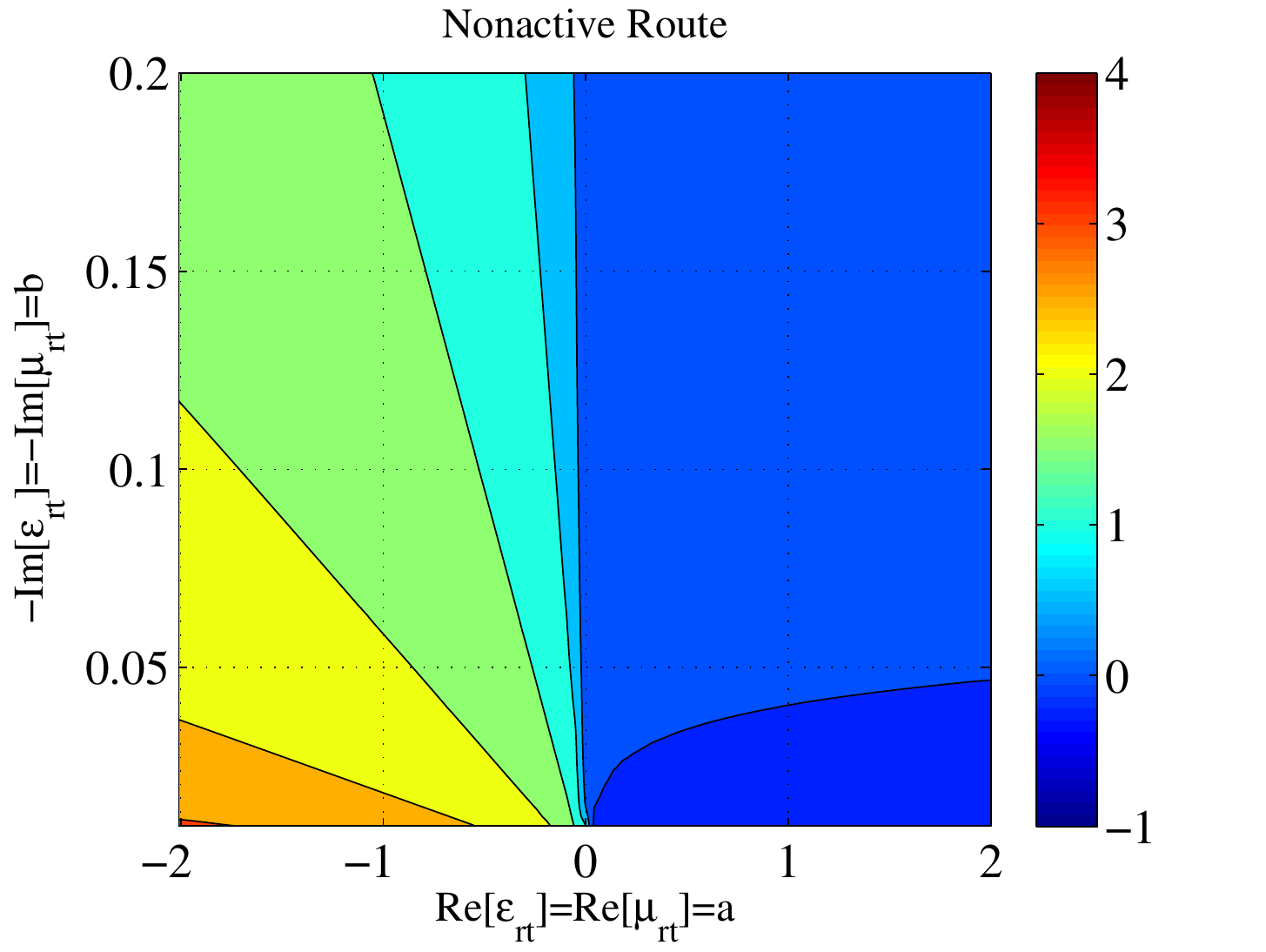}
   \label{fig:Fig5b}}
\subfigure[]{\includegraphics[scale =0.32]{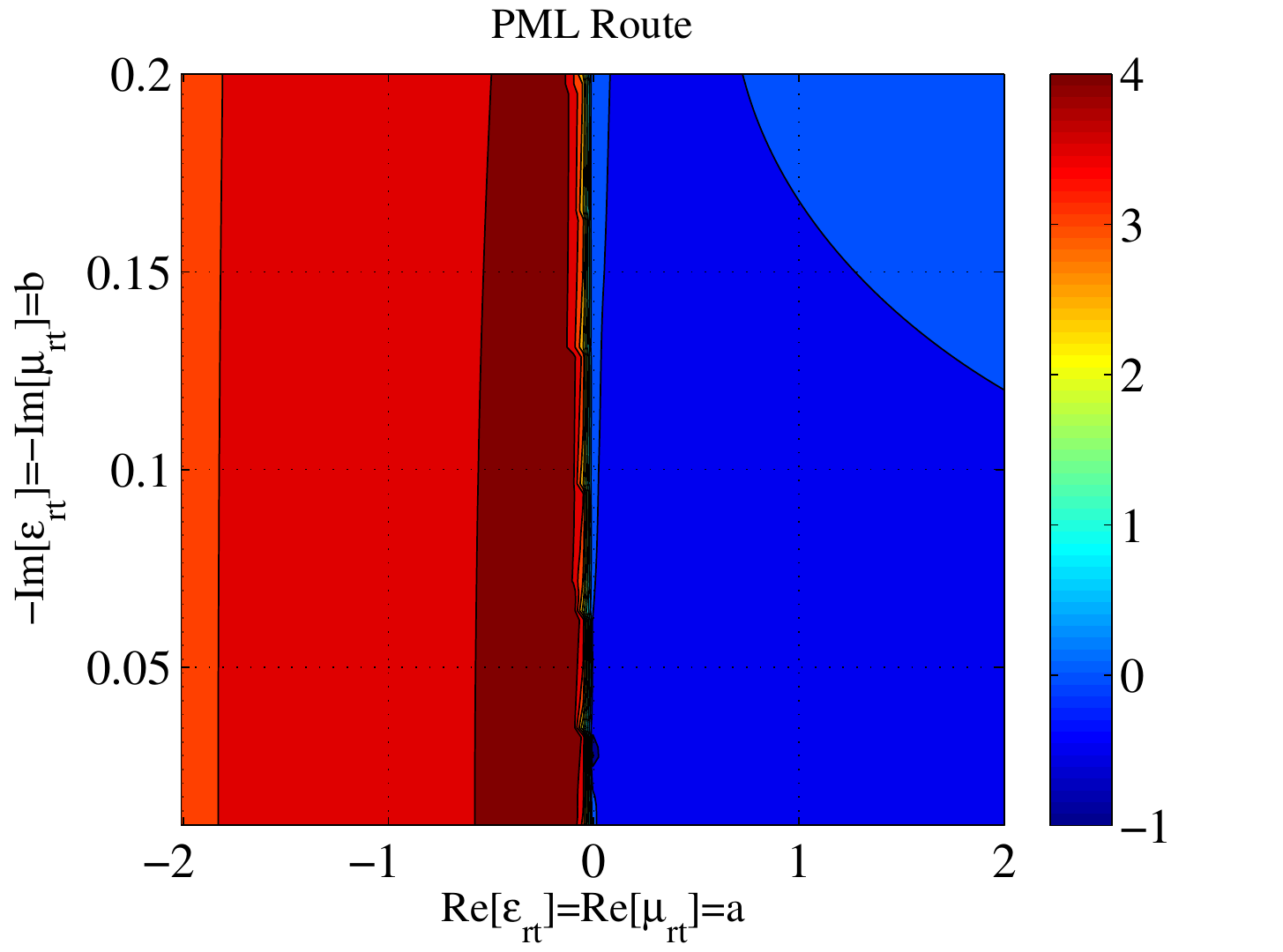}
   \label{fig:Fig5c}}
\caption{Contour plots of the quantity $\log\left(P/P_0\right)$ on the complex plane of the transversal constituent relative parameters $(a=\Re[\varepsilon_{rt}]=\Re[\mu_{rt}], b=-\Im[\varepsilon_{rt}]=-\Im[\mu_{rt}])$ for the three considered routes. (a) The isotropic route $(\varepsilon_{rn}=\varepsilon_{rt}=\mu_{rt}=a-jb)$. (b) The nonactive route $\left(\varepsilon_{rn}=\Re\left[1/\varepsilon_{rt}\right]=\Re\left[1/\mu_{rt}\right]=a/(a^2+b^2)\right)$. (c) The PML route $\left(\varepsilon_{rn}=1/\varepsilon_{rt}-j\delta=\right.$ $\left.1/\mu_{rt}-j\delta=1/(a-jb)-j\delta\right)$. Plot parameters: $k_0L=20$, $k_0g=0.5$, $\theta=90^\circ$, $\delta=0.001$.}
\label{fig:Figs5}
\end{figure}

\begin{figure}[ht]
\centering
\subfigure[]{\includegraphics[scale =0.28]{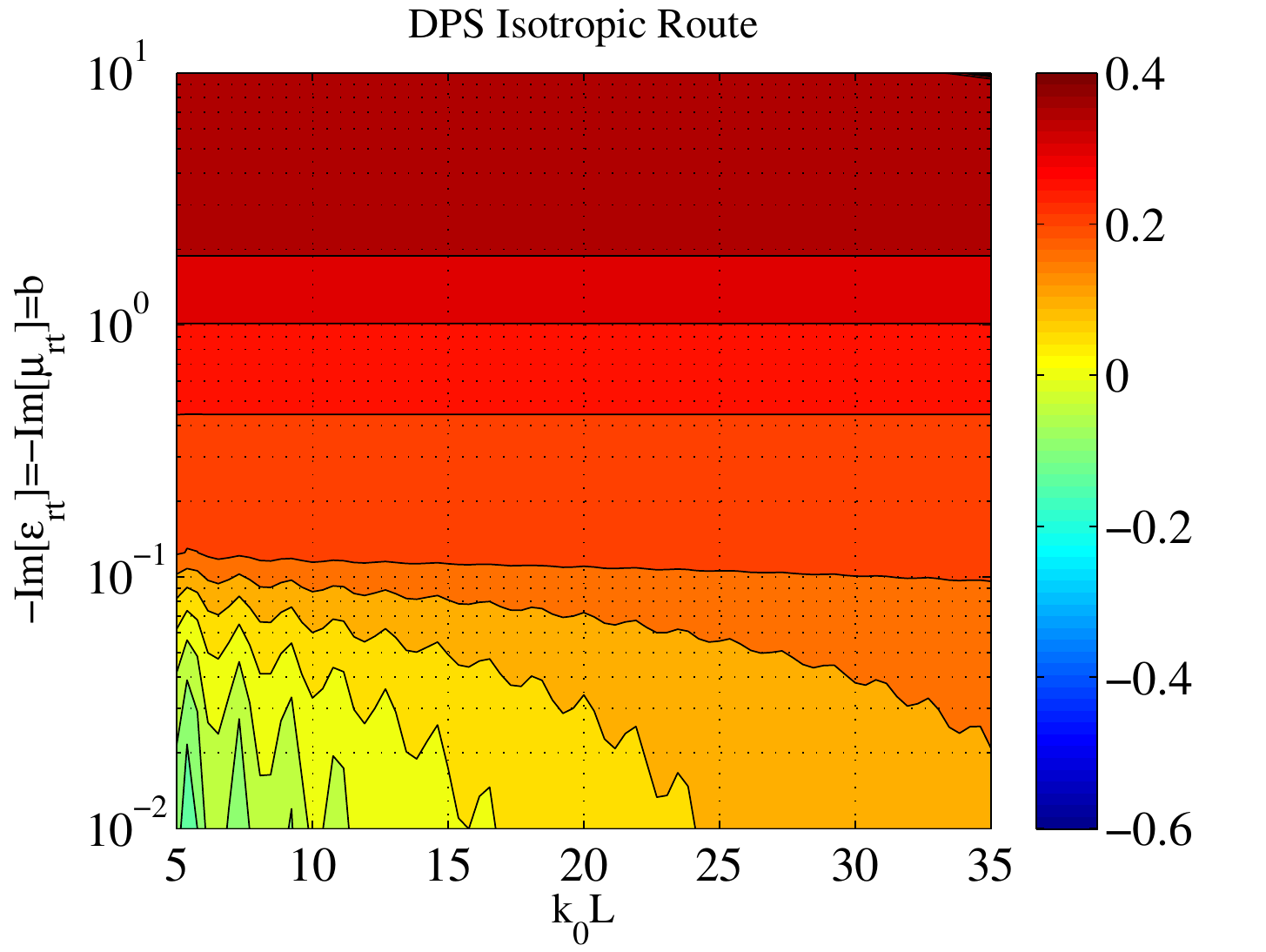}
   \label{fig:Fig6a}}
\subfigure[]{\includegraphics[scale =0.28]{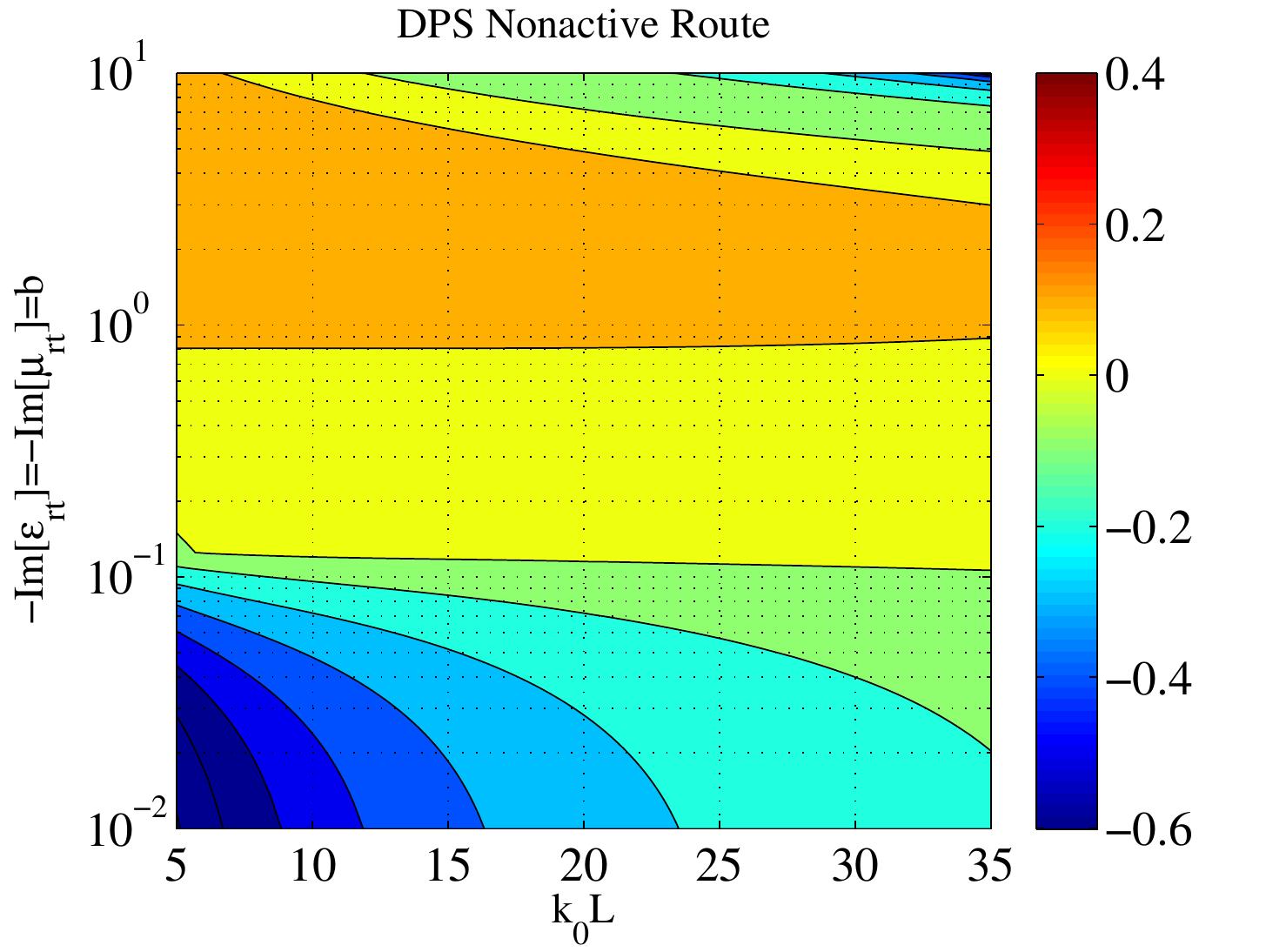}
   \label{fig:Fig6b}}
\subfigure[]{\includegraphics[scale =0.32]{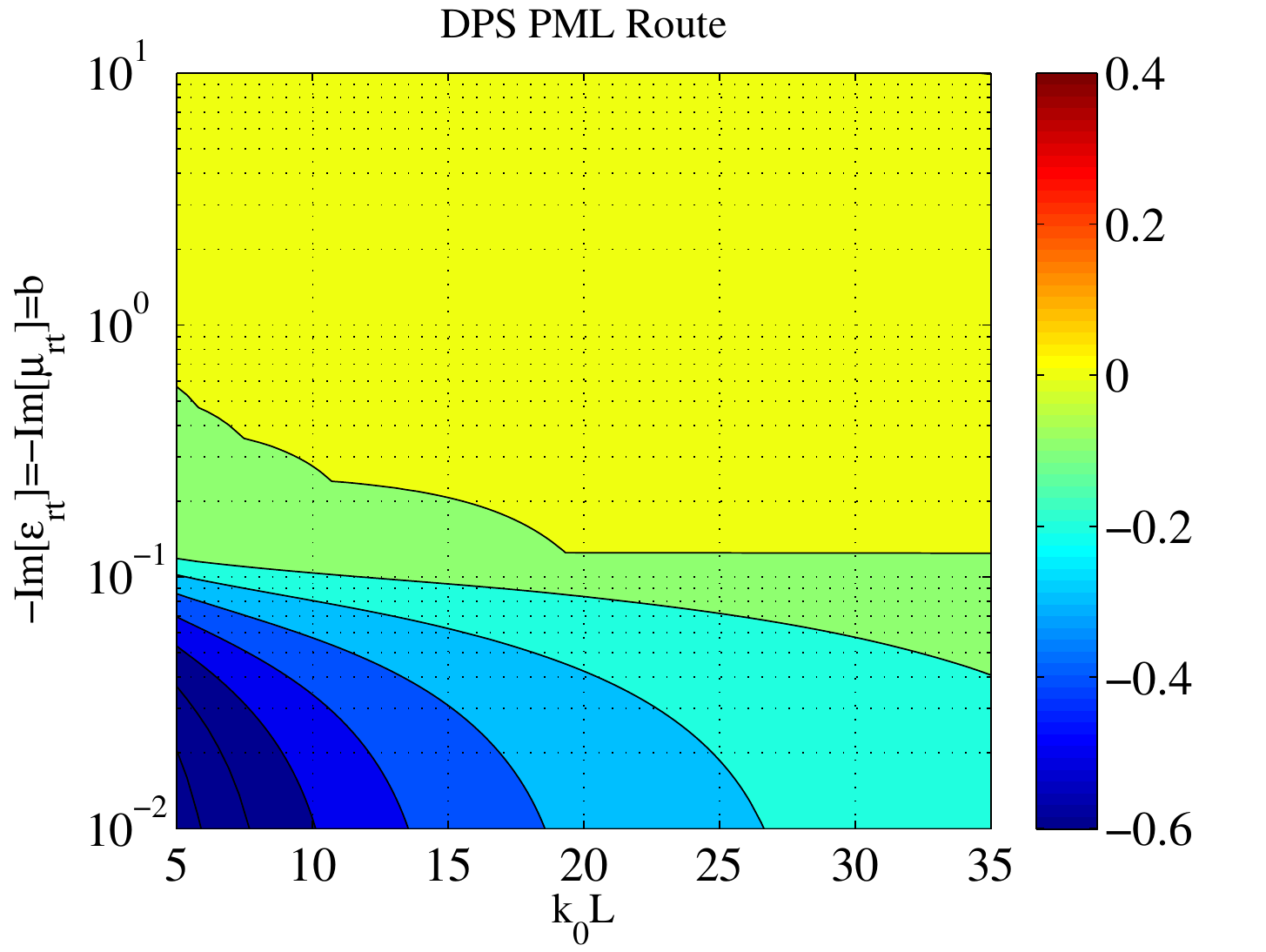}
   \label{fig:Fig6c}}
\caption{Contour plots of the quantity $\log\left(P/P_0\right)$ with respect to the electrical thickness of the grounded slab $k_0L$ and the losses of the transversal constituent parameters $b=-\Im[\varepsilon_{rt}]=-\Im[\mu_{rt}]$ for the three considered DPS routes. (a) The isotropic route $(\varepsilon_{rn}=\varepsilon_{rt}=\mu_{rt}=a-jb)$. (b) The nonactive route $\left(\varepsilon_{rn}=\Re\left[1/\varepsilon_{rt}\right]=\Re\left[1/\mu_{rt}\right]=\Re\left[1/(a-jb)\right]=a/(a^2+b^2)\right)$. (c) The PML route $\left(\varepsilon_{rn}=1/\varepsilon_{rt}-j\delta=\right.$ $\left.1/\mu_{rt}-j\delta=1/(a-jb)-j\delta\right)$. Plot parameters: $a=\Re[\varepsilon_{rt}]=\Re[\mu_{rt}]=2$, $k_0g=0.5$, $\theta=90^\circ$, $\delta=0.001$.}
\label{fig:Figs6DPS}
\end{figure}

\begin{figure}[ht]
\centering
\subfigure[]{\includegraphics[scale =0.28]{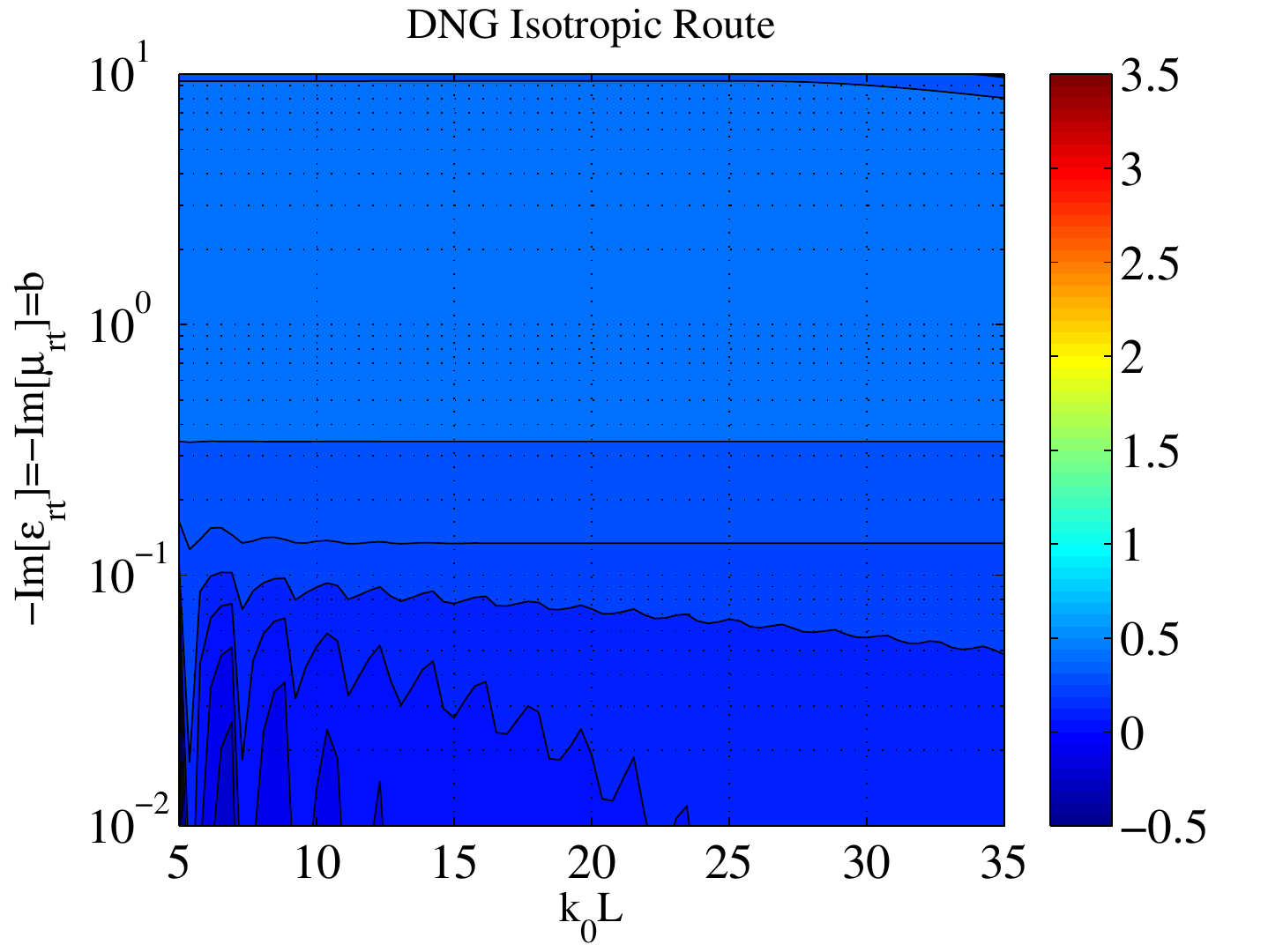}
   \label{fig:Fig6d}}
\subfigure[]{\includegraphics[scale =0.28]{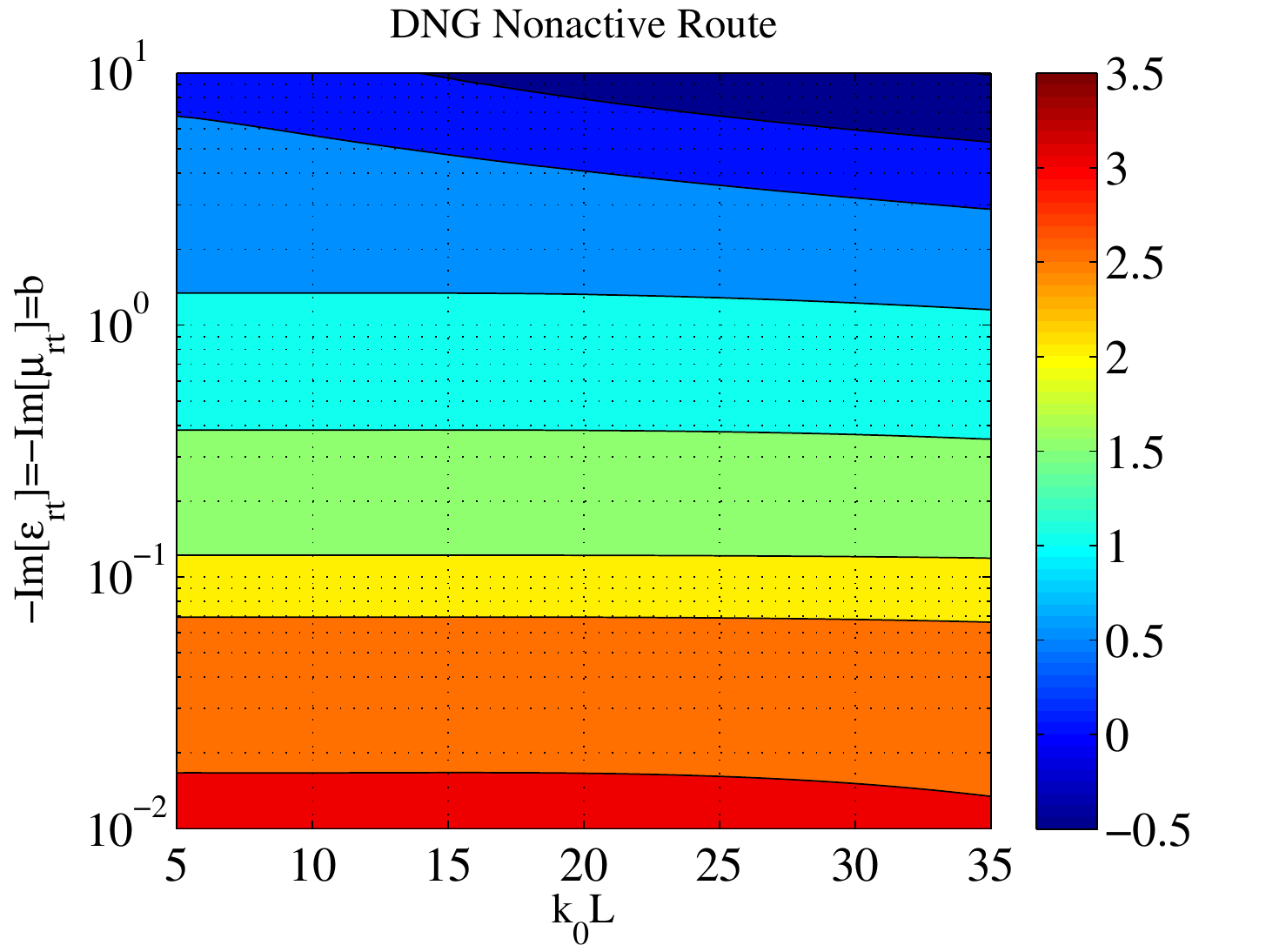}
   \label{fig:Fig6e}}
\subfigure[]{\includegraphics[scale =0.32]{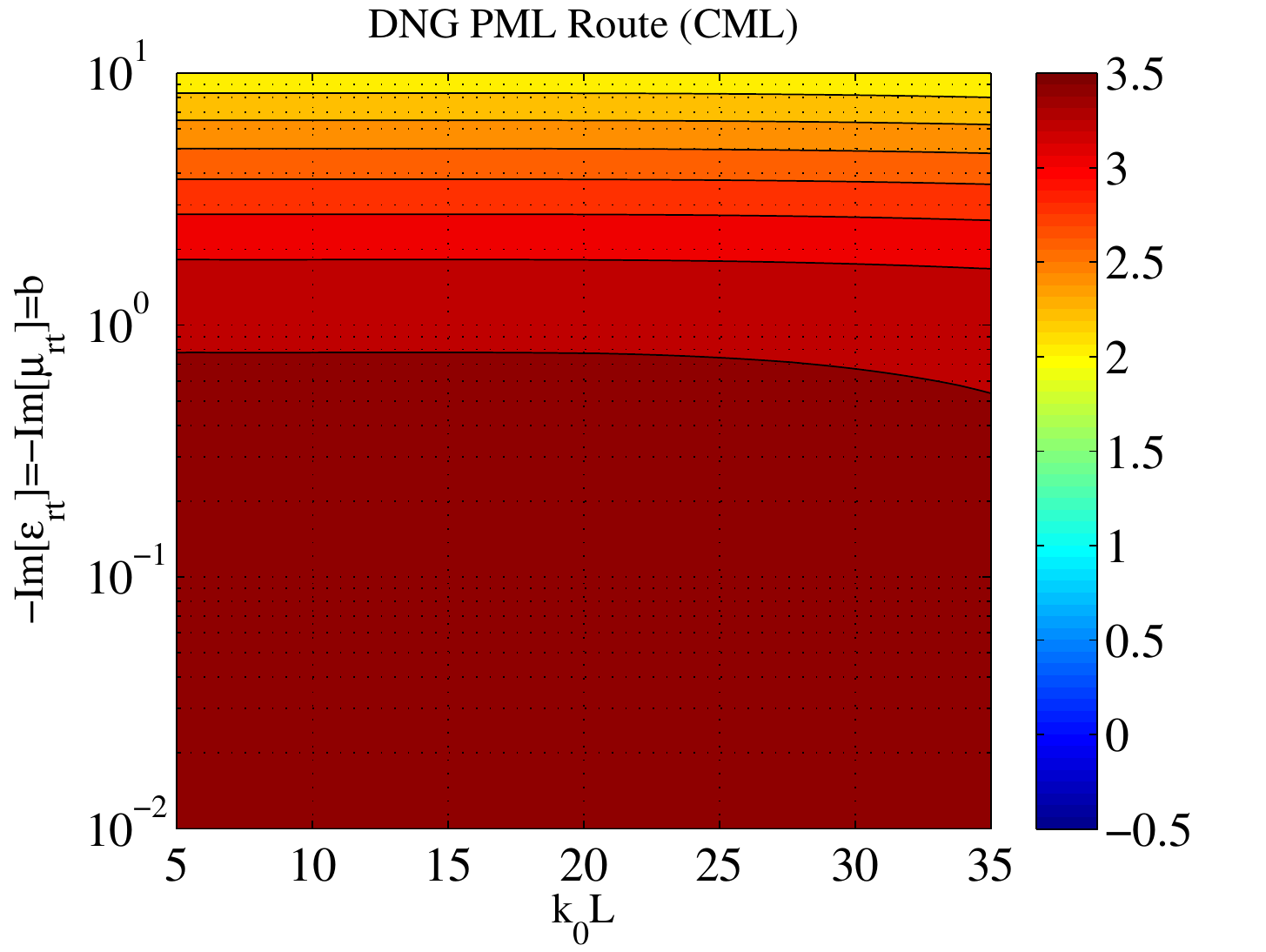}
   \label{fig:Fig6f}}
\caption{Contour plots of the quantity $\log\left(P/P_0\right)$ with respect to the electrical thickness of the grounded slab $k_0L$ and the losses of the transversal constituent parameters $b=-\Im[\varepsilon_{rt}]=-\Im[\mu_{rt}]$ for the three considered DNG routes. (a) The isotropic route $(\varepsilon_{rn}=\varepsilon_{rt}=\mu_{rt}=a-jb)$. (b) The nonactive route $\left(\varepsilon_{rn}=\Re\left[1/\varepsilon_{rt}\right]=\Re\left[1/\mu_{rt}\right]=\Re\left[1/(a-jb)\right]=a/(a^2+b^2)\right)$. (c) The PML route $\left(\varepsilon_{rn}=1/\varepsilon_{rt}-j\delta=\right.$ $\left. 1/\mu_{rt}-j\delta=1/(a-jb)-j\delta\right)$. Plot parameters: $a=\Re[\varepsilon_{rt}]=\Re[\mu_{rt}]=-2$, $k_0g=0.5$, $\theta=90^\circ$, $\delta=0.001$.}
\label{fig:Figs6DNG}
\end{figure}

In Figs.~\ref{fig:Figs5} we represent the absorption enhancement parameter $\log\left(P/P_0\right)$ on the complex permittivity map $(a,b)$ with $\varepsilon_{rt}=\mu_{rt}=a-jb$ for the three considered routes. In the isotropic scenario (Fig.~\ref{fig:Fig5a}), the overall absorption is small while a local maximum is exhibited along the line $a=-1$, which corresponds to the Veselago material \cite{VeselagoMaterial}. This result is expected, because in the absence of losses this medium is conjugate  matched for all modes at this single point \cite{DNG, modeboo}. As far as the nonactive route (Fig.~\ref{fig:Fig5b}) is concerned, we observe much better performance compared to the isotropic route: the power extracted from the source is large in a wide range of material parameters, and it is monotonically growing when $|a|$ increases along the negative semi-axis ($a<0$). Most importantly, in Fig.~\ref{fig:Fig5c}, where the PML case is examined, we clearly notice a huge change in the absorbed power $P$ from double-positive $(a>0)$ to double-negative $(a<0)$ cases, as expected from (\ref{EvanescentAbsorbedPower}). In fact, extremely high enhancement of absorption is achieved for the CML case ($P\cong10^4P_0$) which means that the structure is sucking all the power (for every single mode, propagated or evanescent) from the source.

In Figs.~\ref{fig:Figs6DPS} and \ref{fig:Figs6DNG} we show the variation of $\log\left(P/P_0\right)$ with respect to the electrical thickness of the grounded slab $k_0L$ and the loss factors $b$ of its relative constituent parameters for the three routes  (isotropic, nonactive, and lossy PML) for different signs of $a$ (Figs. \ref{fig:Figs6DPS} for $a>0$ and Figs.~\ref{fig:Figs6DNG} for $a<0$). In the isotropic scenario of the double-positive case (Fig.~\ref{fig:Fig6a}), we observe resonances when $k_0L$ is moderate and the reflected fields from $x=L$ are quite strong to interfere with the incident waves, while the absorption $P$ is positively related to the loss parameter $b$. When it comes to the DPS nonactive scenario (Fig.~\ref{fig:Fig6b}), we have very small absorption for $b\rightarrow 0$ (as physically anticipated), but there is an optimal range of $b$, for each fixed $k_0L$ across which the absorbed power gets maximized. It is also remarkable that the DPS PML configuration (Fig.~\ref{fig:Fig6c}) performs worse than the other two in terms of absorption; furthermore, note that $P$ is not substantially dependent on losses $b$, as is clear from (\ref{eq:PDPSGevanescentDivergent}), (\ref{eq:PDPSGevanescentConvergent}). As far as the double-negative configurations are concerned, we remark that the behavior of the structure following the isotropic route (Fig.~\ref{fig:Fig6d}) is similar to that of the corresponding double-positive configuration (Fig.~\ref{fig:Fig6a}). On the contrary, the nonactive realization (Fig.~\ref{fig:Fig6e}) does well for small losses $b$ and its absorbing efficiency deteriorates with increasing $b$. Finally, the CML structure (Fig.~\ref{fig:Fig6f}) absorbs, on the average, extremely high power $P$, as predicted by (\ref{eq:PDNGevanescent}), and the harvested power is practically not fluctuating with the electrical thickness $k_0L$, because the evanescent mode absorption is due to resonant surface modes.

\begin{figure}[ht]
\centering
\subfigure[]{\includegraphics[scale =0.5]{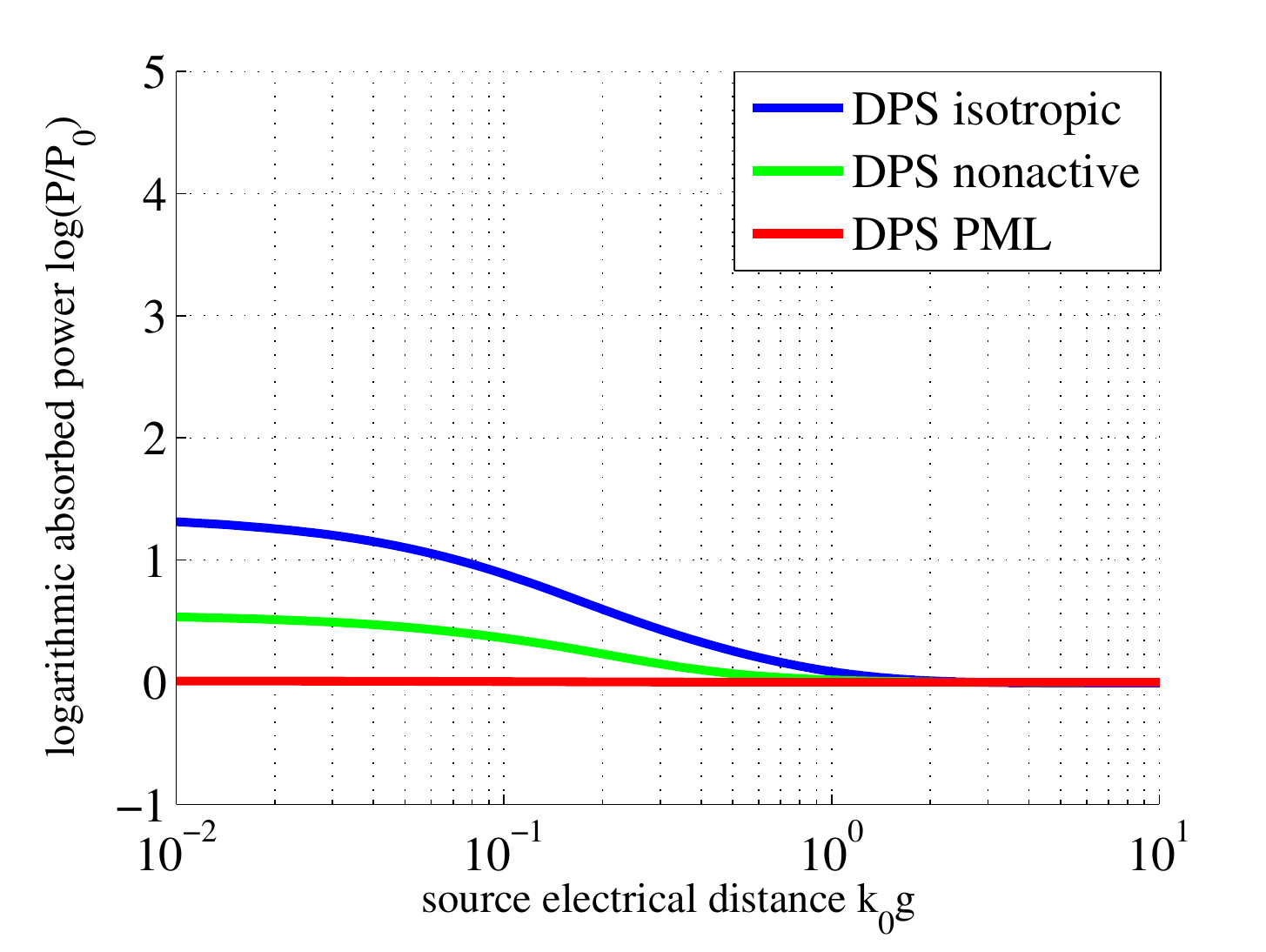}
   \label{fig:Fig7a}}
\subfigure[]{\includegraphics[scale =0.5]{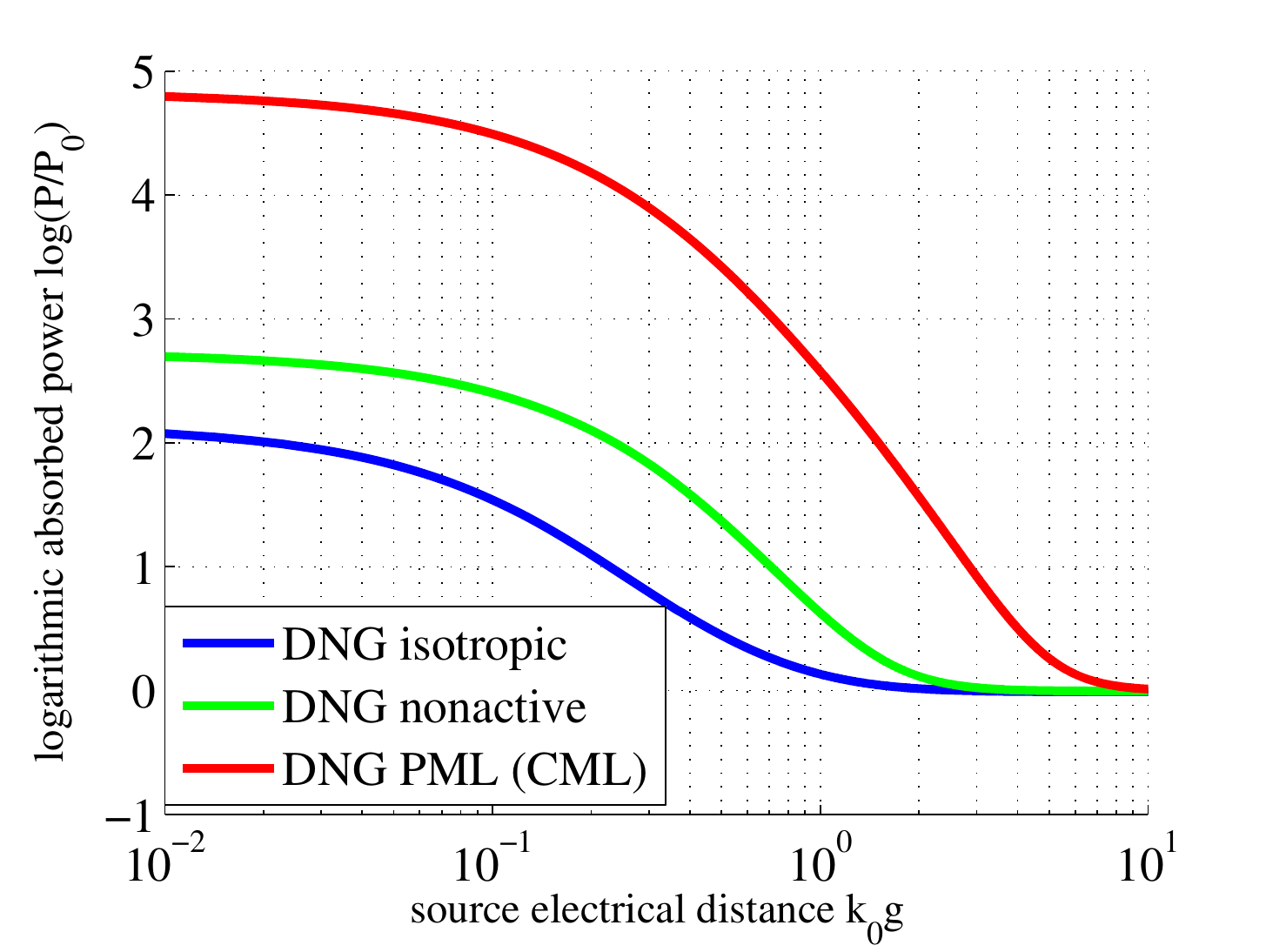}
   \label{fig:Fig7b}}
\caption{The quantity $\log\left(P/P_0\right)$ as function of the electrical distance of the source $k_0g$ from the boundary for the three considered routes. (a) Double-positive cases $a=\Re[\varepsilon_{rt}]=\Re[\mu_{rt}]=2$. (b) Double-negative cases $a=\Re[\varepsilon_{rt}]=\Re[\mu_{rt}]=-2$. Plot parameters: $b=-\Im[\varepsilon_{rt}]=-\Im[\mu_{rt}]=0.5$, $k_0L=20$, $\theta=90^\circ$, $\delta=0.001$.}
\label{fig:Figs7}
\end{figure}

In Figs.~\ref{fig:Figs7} we represent the logarithm of the relative absorbed power $\log\left(P/P_0\right)$ with respect to the electrical distance from the source to the boundary $k_0g$. All the three scenarios are considered for double-positive (Fig.~\ref{fig:Fig7a}) and double-negative (Fig.~\ref{fig:Fig7b}) configurations. One can notice that the DPS PML absorbs power $P_0$ only from propagating waves since $a>0$ and $\delta\rightarrow 0$, as suggested by (\ref{eq:PDPSGevanescentDivergent}), (\ref{eq:PDPSGevanescentConvergent}); in other words, PML is outperformed by the other two scenarios in the double-positive paradigm. Again, we remark a similar response of the isotropic structure, regardless of the source location $k_0g$ (the blue curves in the two figures), while the nonactive configuration performs always in-between the other two scenarios (which are switching positions when changing the sign of $a$). In Fig.~\ref{fig:Fig7b} we observe huge absorption for the DNG CML case at moderate distances to the source ($P\cong10^5P_0$), as expected from (\ref{eq:PDNGevanescent}). Finally, when $k_0g\rightarrow+\infty$, according to all the scenarios, the device tends to absorb power $P=P_0$ which indicates that only propagating waves survive and excite the boundary. The magnitude is the same since, due to the position of the source ($\theta=90^\circ$), the illumination is almost normally incident on the grounded slab and only the transversal relative parameters (the same for all the scenarios, $\varepsilon_{rt}=\mu_{rt}$) are activated.

\begin{figure}[ht]
\centering
\subfigure[]{\includegraphics[scale =0.5]{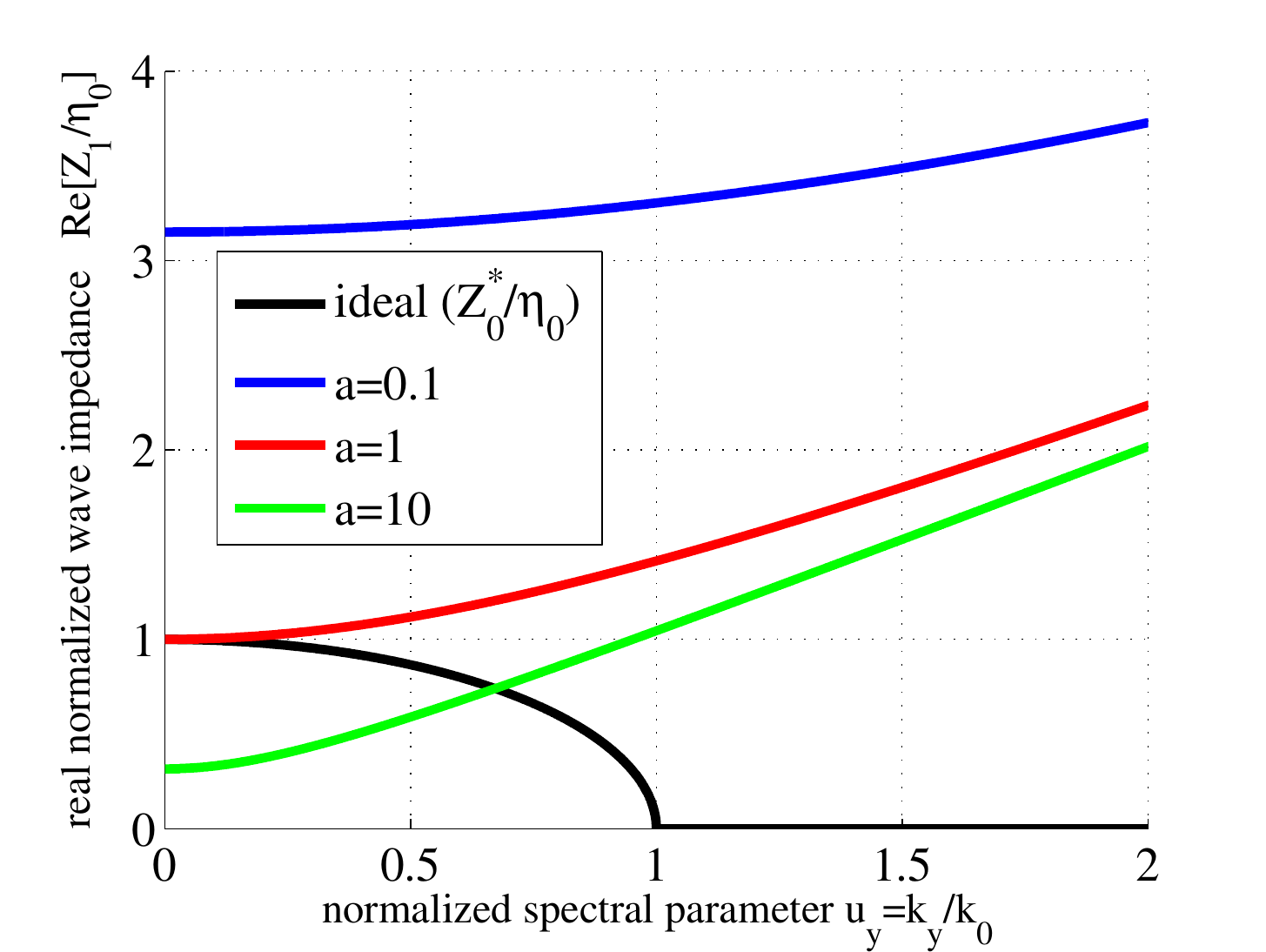}
   \label{fig:Fig74a}}
\subfigure[]{\includegraphics[scale =0.5]{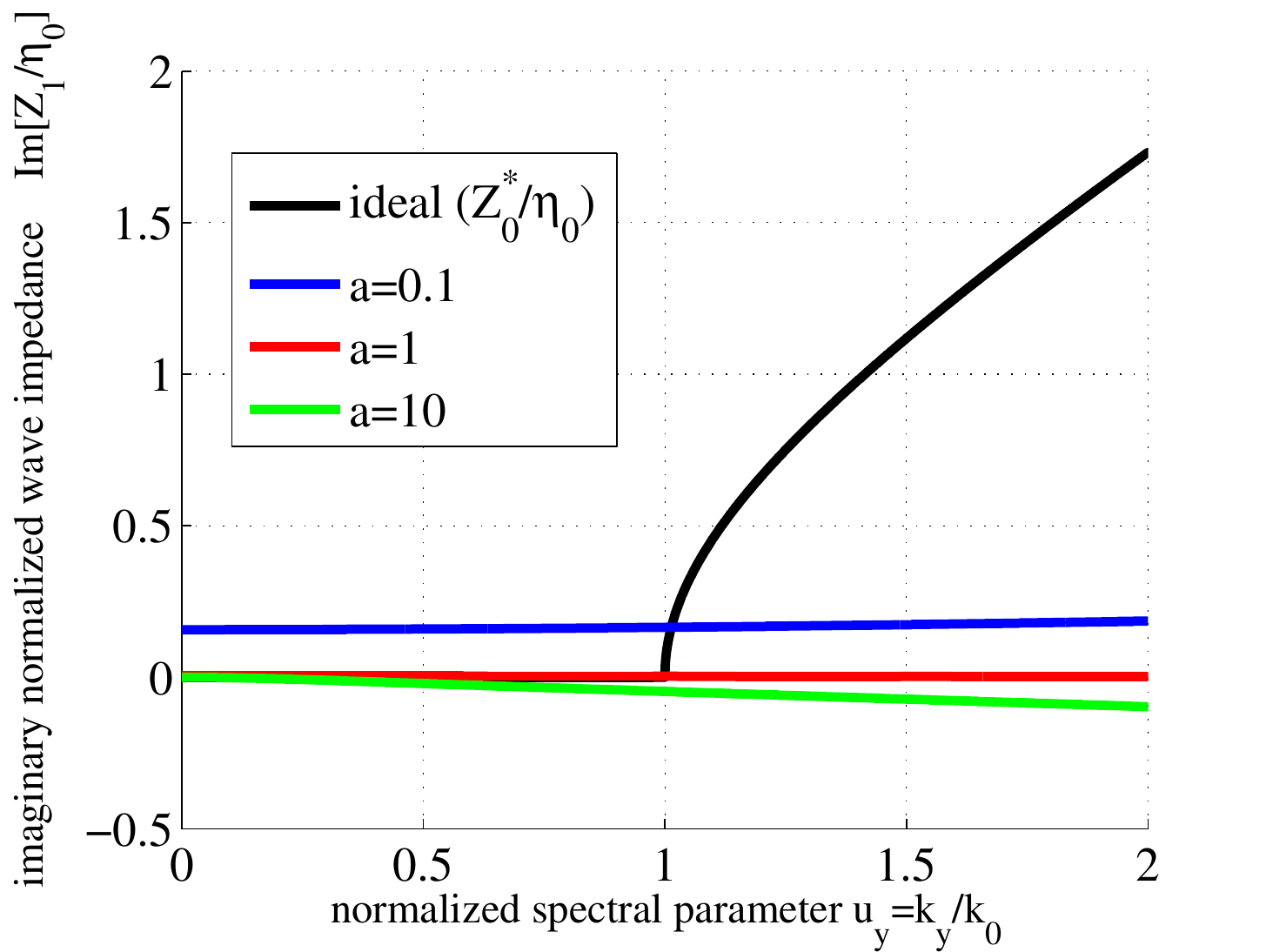}
   \label{fig:Fig74b}}
\caption{The variation of: (a) the real parts and (b) the imaginary parts of the normalized TM impedance $Z_1/\eta_0$ for a uniaxial hyperbolic medium with $\varepsilon_{rt}=a-jb, \mu_{rt}=1, \varepsilon_{rn}=-1/a-jb$ as function of the normalized spectral parameter $u_y$ for various $a$. The ideal for conjugate matching impedance $Z_0^*/\eta_0$ is also shown.}
\label{fig:Figs74}
\end{figure}

\begin{figure}[ht]
\centering
\subfigure[]{\includegraphics[scale =0.28]{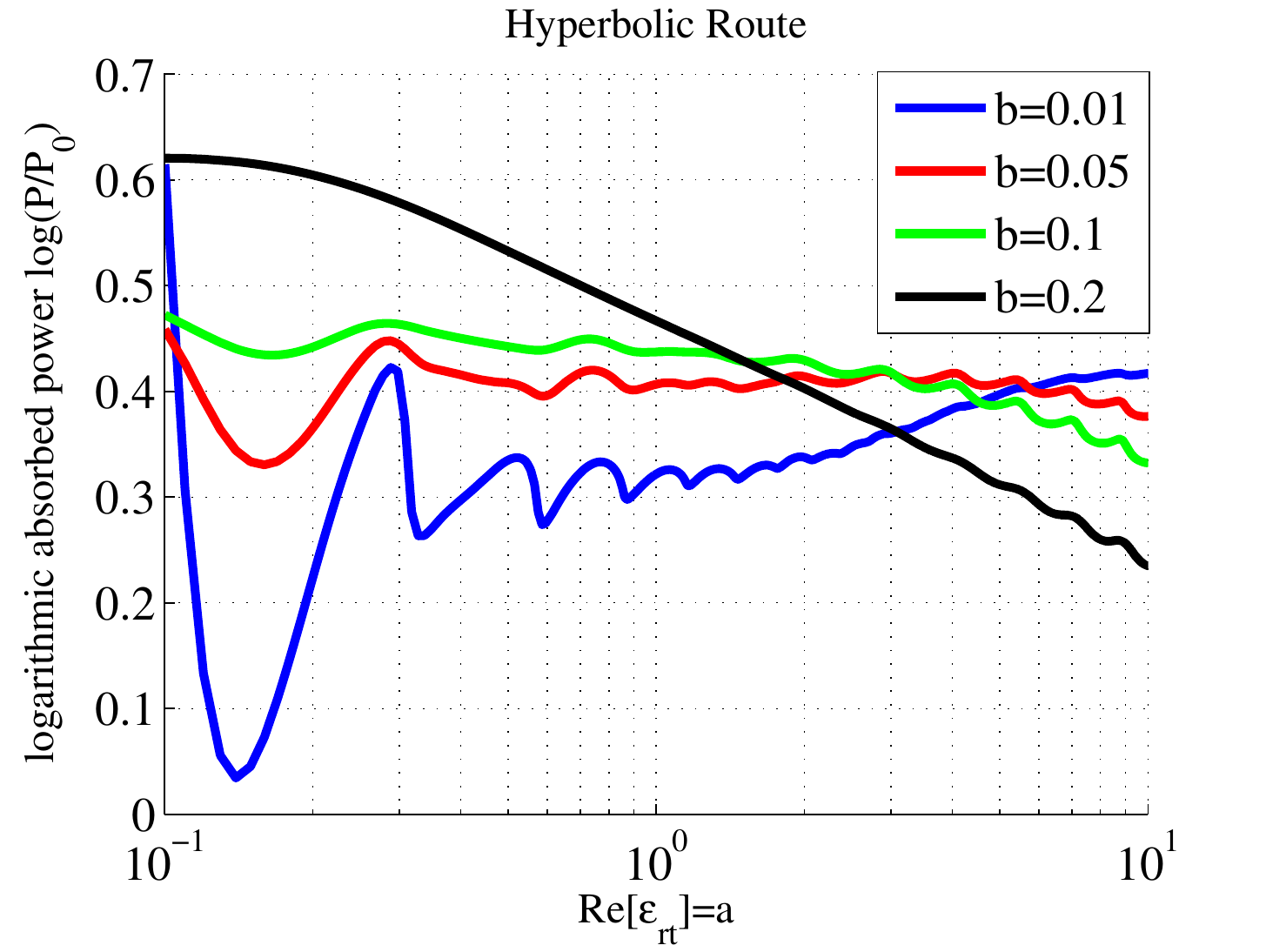}
   \label{fig:Fig75a}}
\subfigure[]{\includegraphics[scale =0.28]{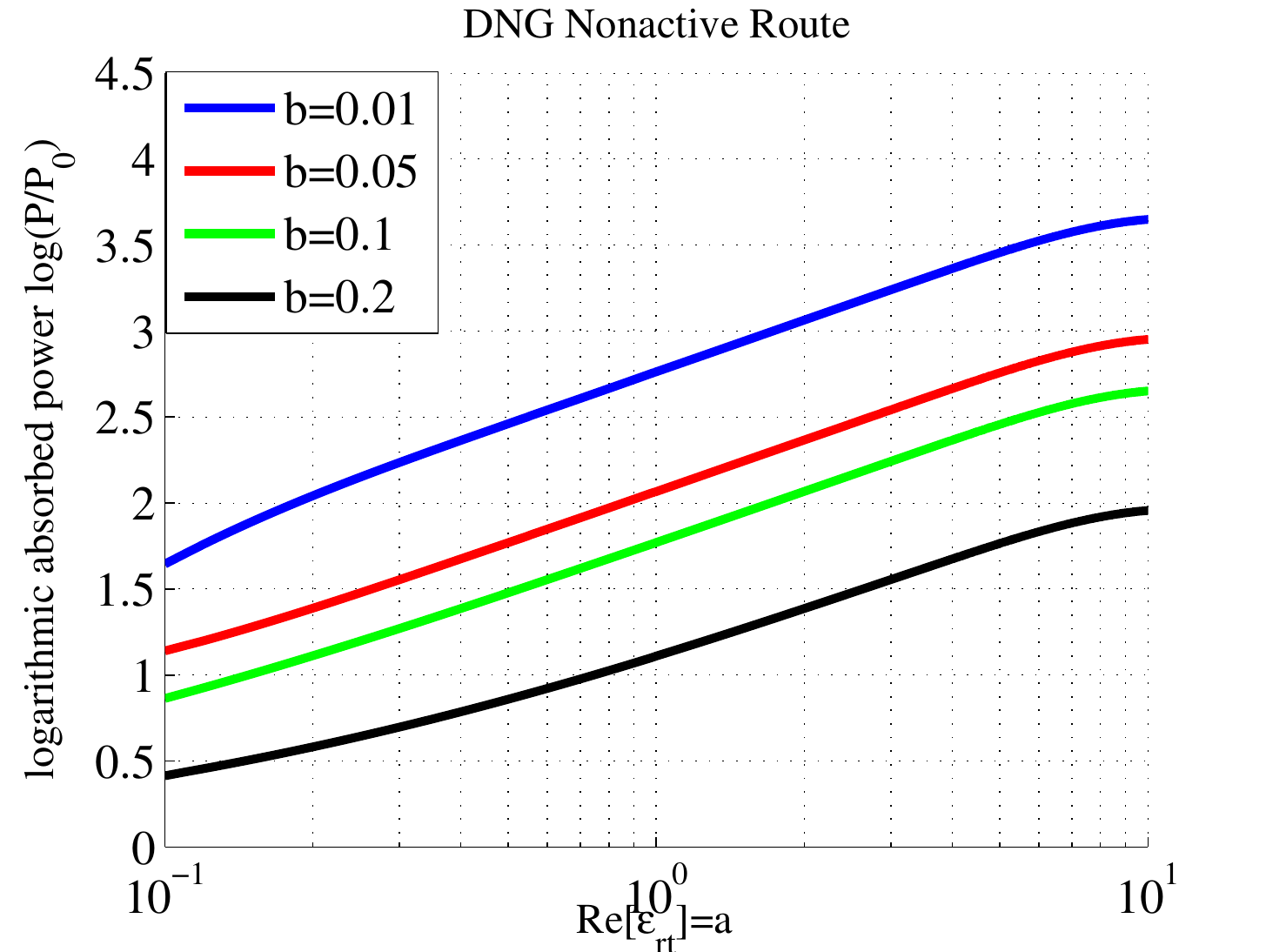}
   \label{fig:Fig75b}}
\subfigure[]{\includegraphics[scale =0.32]{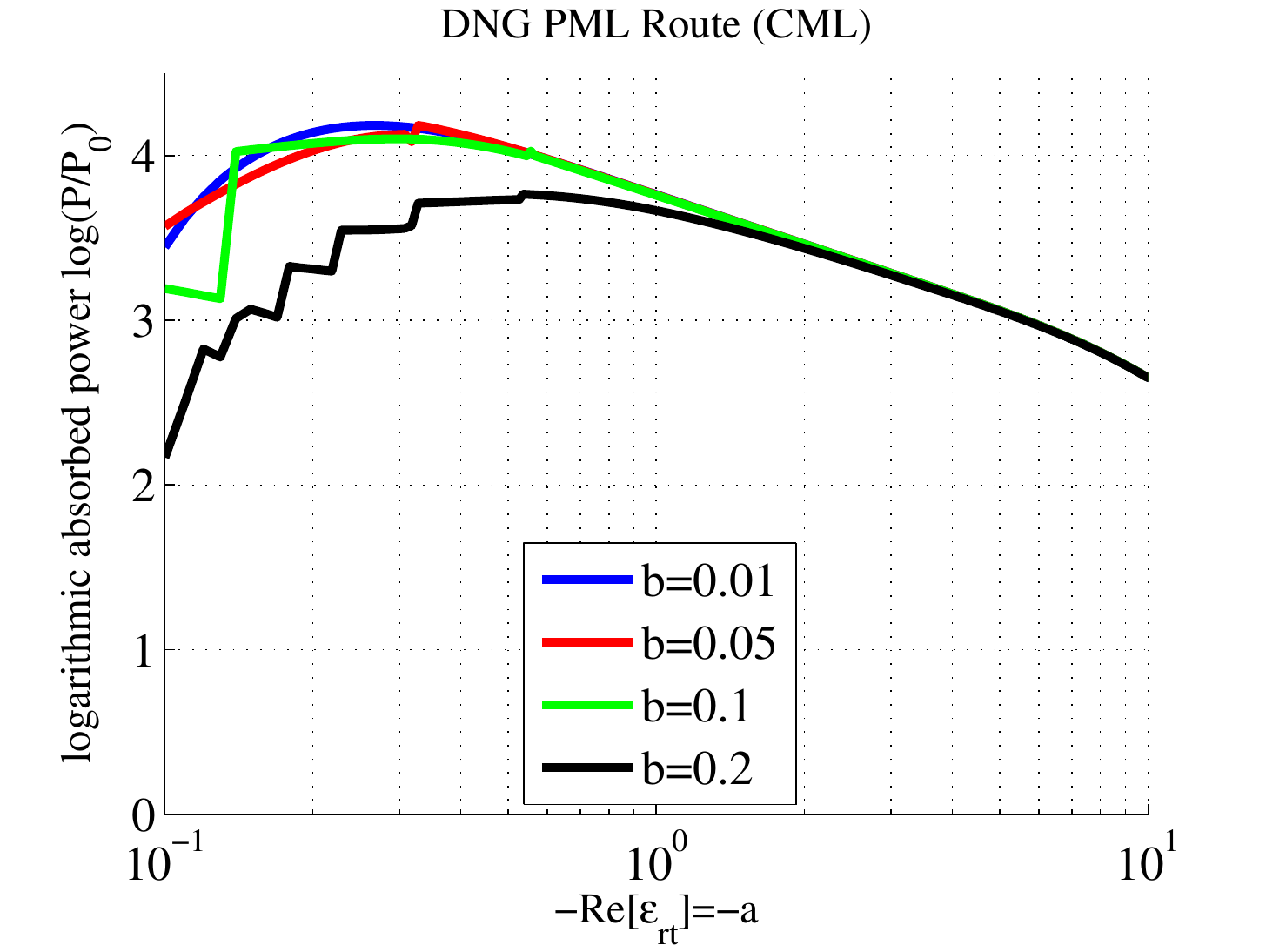}
   \label{fig:Fig75c}}
\caption{The quantity $\log\left(P/P_0\right)$ as function of the transversal permittivity $a$ for several losses $b$, evaluated for: (a) the hyperbolic case $\varepsilon_{rt}=a-jb, \mu_{rt}=1, \varepsilon_{rn}=-1/a-jb$ with $a>0$. (b) the nonactive route case $\varepsilon_{rt}=\mu_{rt}=a-jb, \varepsilon_{rn}=\Re\left[1/(a-jb)\right]=a/(a^2+b^2)$ with $a<0$. (c) the PML route case $\varepsilon_{rt}=\mu_{rt}=a-jb, \varepsilon_{rn}=1/(a-jb)-j\delta$ with $a<0$. Plot parameters: $k_0L=10$, $k_0g$=0.5, $\theta=90^\circ$, $\delta=0.001$.}
\label{fig:Figs75}
\end{figure}

To conclude this section we note that although we have considered only one (TM) polarization, all the results are general and hold also for the TE polarization, which can be studied in the same way.

\subsection{Comparison with Hyperbolic Media}

Probably the only known alternative approach to enhance absorption of evanescent fields in large lossy bodies is the use of \emph{hyperbolic} media \cite{hyperbolic_review, HyperbolicBlackbody, MetamaterialSpontaneousEmission, hyp_shalaev, DipoleLineSource}. Due to the transformation of evanescent waves to propagating ones in hyperbolic media, substantial absorption enhancement has  been observed \cite{MetamaterialSpontaneousEmission,hyp_shalaev}. Therefore, it would be meaningful to compare our results with those obtained when utilizing hyperbolic materials. In hyperbolic media, the real parts of different eigenvalues of the permittivity tensor have opposite signs. Thus, we consider the same structure of Fig.~\ref{fig:Fig4a} with $\varepsilon_{rt}=a-jb$, $\mu_{rt}=1$ and $\varepsilon_{rn}=-\frac{1}{a-jb}$ for $a>0$ and small $b>0$. We assume that $\Re[\varepsilon_{rt}]\Re[\varepsilon_{rn}]=-1$, because the maximal absorption is achieved  under this assumption \cite{HyperbolicBlackbody}. In Figs.~\ref{fig:Figs74} we present the relative wave impedance $Z_1/\eta_0$ (the real and imaginary parts) of the hyperbolic material as functions of the normalized spectral parameter $u_y$ for several $a$. It is apparent that along the entire $u_y$ axis, the differences from the ideal $Z_0^*/\eta_0$ are substantial, which means that the absorption would be dramatically smaller than in our conjugate-matching scenarios. This expectation is confirmed by the results shown in Figs.~\ref{fig:Figs75}, where the logarithm of the absorbed power $\log\left(P/P_0\right)$ is presented as a function of the real part of the permittivity $a$ for various loss factors $b$. In the hyperbolic scenario (Fig.~\ref{fig:Fig75a}), the power absorption is, on the average, two orders of magnitude smaller as compared to the absorbed power along the DNG nonactive (Fig.~\ref{fig:Fig75b}) or the PML (Fig.~\ref{fig:Fig75c}) routes. Furthermore, one can point out that for large $|a|$ ($a<0$ for Figs.~\ref{fig:Fig75b} and \ref{fig:Fig75c}) the nonactive route gives better outcomes than the PML route. This property follows from the fact that we have assumed a constant additional loss factor $\delta$ for the PML route. In particular, for $|a|\rightarrow+\infty$, the normal permittivity component of the nonactive material tends to zero (which is the ideal value in the PML route case, corresponding to conjugate matching), but for the PML route, we have $\varepsilon_{rn}\rightarrow -j\delta$. In other words, when $|a|$ is increasing without limit while $\delta$ and $b$ remain fixed, the CML response to evanescent excitations appears more lossy than the response of our nonactive device.

\section{Finite Cylindrical Configuration}
\label{sec:Cylindrical}

\subsection{Conjugate Matched Cylinder via Coordinate Transformation}

Let us next study possibilities for the use of the double-negative conjugate matched materials to create ideal absorbing bodies of \emph{finite} sizes (in cross sections normal to $\hat{\textbf{z}}$). In contrast to an infinite planar surface, studied above, finite-sized bodies can in principle absorb infinite power carried by a single propagating plane wave, generated by sources at infinity \cite{Bobr,unlimited_antenna,SphericalPaper}. A conceptual example of a spherical body having these properties has been described \cite{SphericalPaper}, where the sphere was filled with a locally isotropic low-loss double-negative material. As discussed above, double-negative materials can be ideally conjugate matched to free space, but only in the limit of negligible loss factors. Here we study a cylindrical configuration and explore additional design possibilities offered by double-negative perfectly matched layers, which have more free parameters to tune.

\begin{figure}[ht]
\centering
\subfigure[]{\includegraphics[scale =0.27]{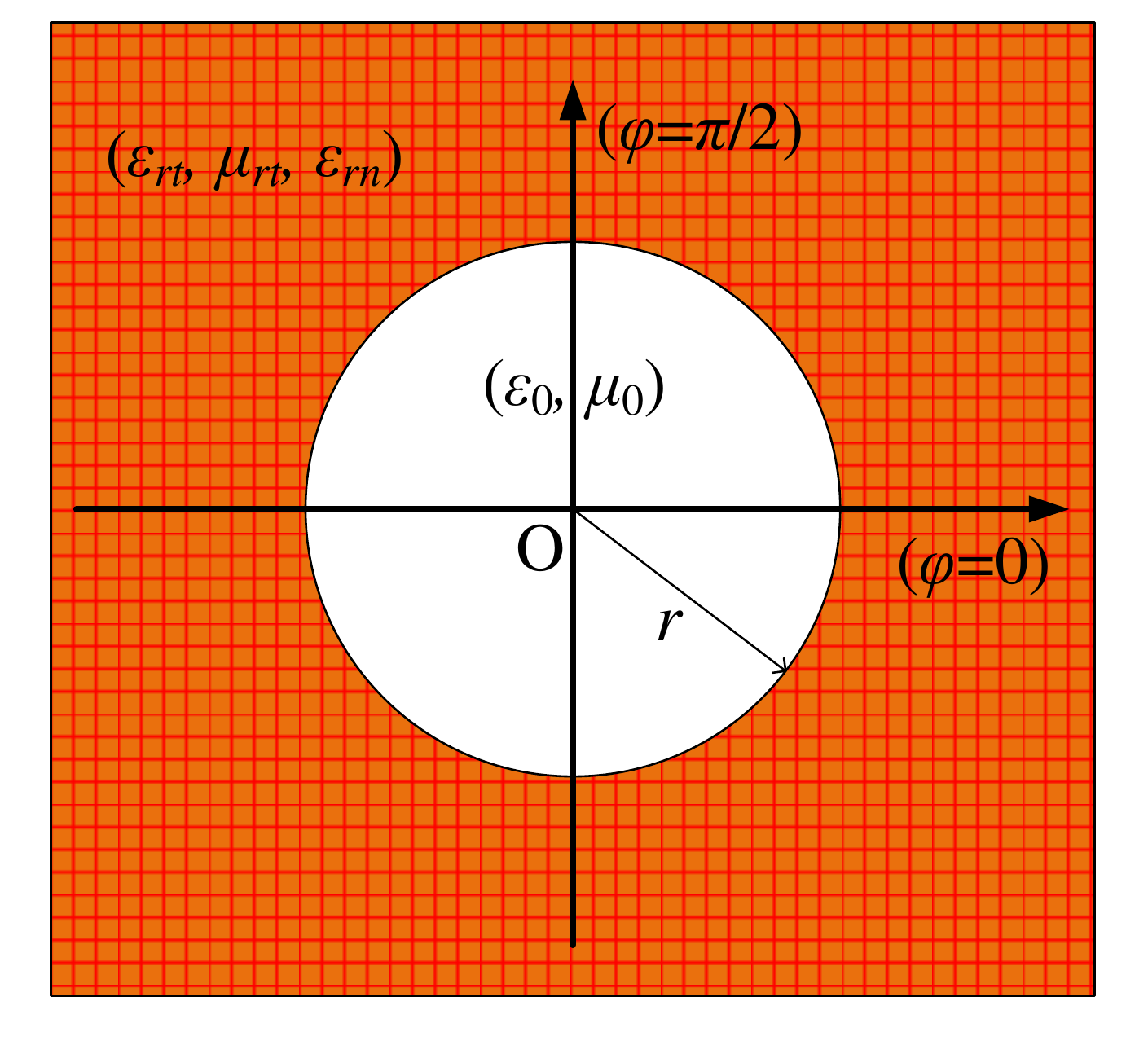}
   \label{fig:Fig8a}}
\subfigure[]{\includegraphics[scale =0.27]{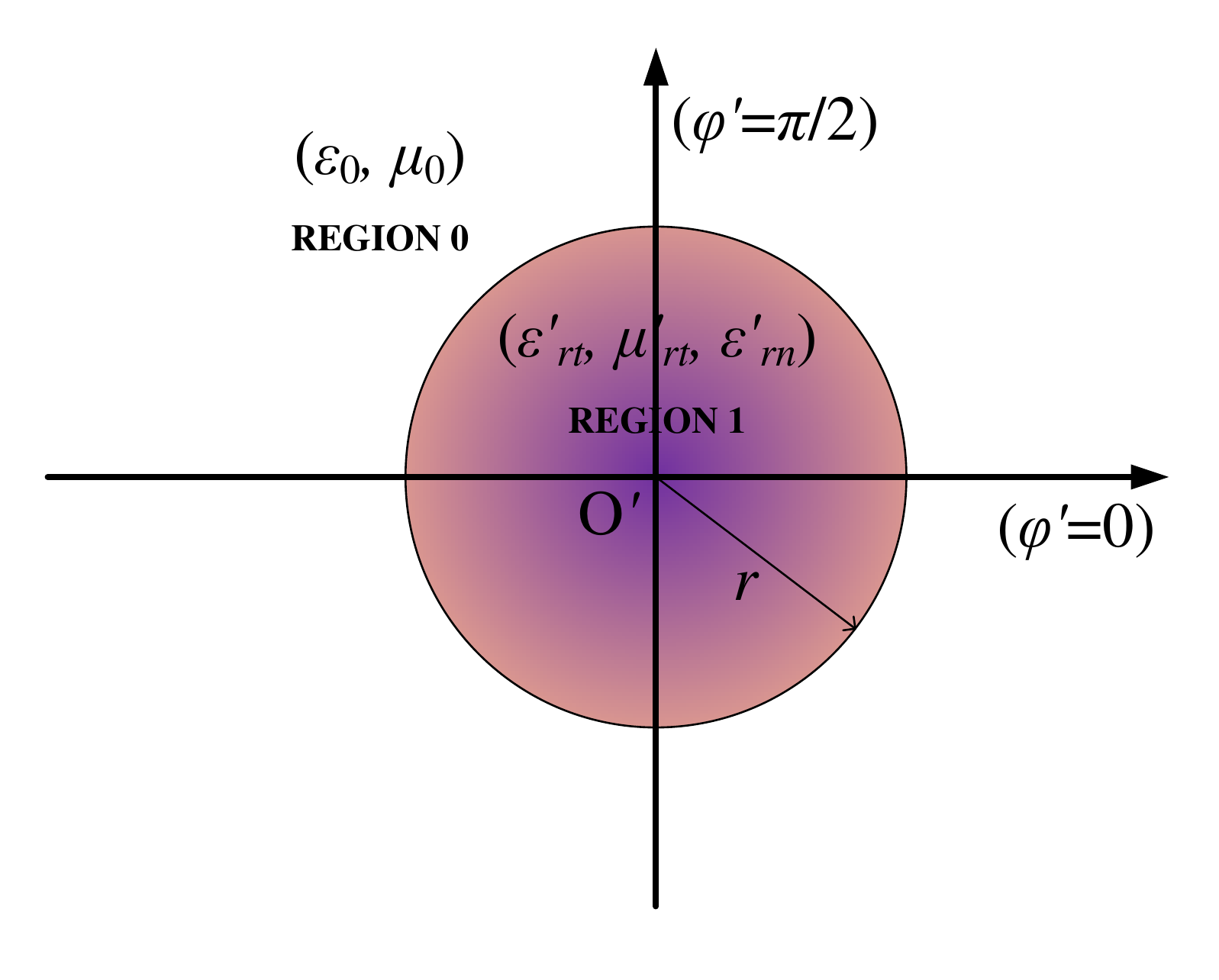}
   \label{fig:Fig8b}}
\caption{(a) Complementary cylindrical infinitely thick mantle with the internal boundary $\rho=r$ filled with the conventional PML relative constituent parameters $\left(\varepsilon_0[\varepsilon_r], \mu_0[\mu_r]\right)$ located in the unprimed coordinate system $(\rho,\varphi,z)$. (b) Cylindrical core with the external boundary $\rho'=r$ filled with a material having transformed constituent parameters $\left(\varepsilon_0[\varepsilon^\prime_r], \mu_0[\mu^\prime_r]\right)$ of (\ref{eq:TransformedParameters}), located in the primed coordinate system $(\rho',\varphi',z')$.}
\label{fig:Figs8}
\end{figure}

With the goal of finding the material parameters of a cylindrical CML, we adopt an already proposed approach \cite{SphericalPaper} and consider first a complementary infinitely long cylindrical shell filling the space from $\rho=r$ to $\rho=\infty$ (see Fig.~\ref{fig:Fig8a}, where the used cylindrical coordinate system $(\rho,\varphi,z)$ is also defined). We first assume that the shell is filled with a medium characterized by the following, more general (compared to the PML case) permittivity and permeability dyadics in cylindrical coordinates: [$\varepsilon_r] = [\mu_r] = \textnormal{diag} (\varepsilon_{r\rho},\varepsilon_{r\varphi},\varepsilon_{rz})$.

In order to obtain the corresponding material parameters for a cylinder with a finite radius $r$, we map the outer space of the cylindrical volume $(\rho>r)$ into the internal vacuum hole $(\rho<r)$ and vice versa, as shown in Fig. \ref{fig:Fig8b}. This mapping can be done using the concept of transformation optics \cite{Milton, TransformationOptics}, and it has been already performed in the analogous case of isotropic media and the spherical shape of the conjugate matched body \cite{SphericalPaper}. More specifically, we can use the following coordinate transformation from the unprimed $(\rho,\varphi,z)$ to the primed $(\rho',\varphi',z')$ coordinates:
\begin{equation}
\rho^{\prime}=\frac{r^{n+1}}{\rho^n}, ~~~~~ \varphi^{\prime}=\varphi, ~~~~~ z^{\prime}=-z,
\label{eq:CoordinateTransformation}
\end{equation}
where $n$ can be any positive real number. Note that the sign of the $z$-component changes in the transformation. As $\rho'$ is inversely proportional to $\rho$, the corresponding coordinate axes have opposite directions. Therefore, in order to map a right-handed coordinate system to a right-handed coordinate system, we also change the direction of either the $\hat{\boldsymbol\varphi}$ or $\hat{\textbf{z}}$ axes.

The material parameters transform according to the well-known coordinate transformation equations \cite{TransformationOptics}:
\begin{equation}
[\varepsilon^{\prime}_r] =\frac{\textbf{A}\cdot[\varepsilon_r]\cdot\textbf{A}^T}{\det(\textbf{A})}, \ \ \
[\mu^{\prime}_r]         =\frac{\textbf{A}\cdot[\mu_r]     \cdot\textbf{A}^T}{\det(\textbf{A})},
\label{eq:TransformedParameters}
\end{equation}
where $\star^T$ denotes the transpose of a matrix, $[\varepsilon_r]$ and $[\mu_r]$ are the original material parameter matrices (i.e., the material parameters of the infinite mantle at $\rho>r$), $[\varepsilon^{\prime}_r]$ and $[\mu^{\prime}_r]$ are the transformed material parameter matrices (i.e., the material parameters of the finite core, $\rho<r$), and $\textbf{A}$ is the Jacobian matrix of the transformation. The Jacobian matrix in this particular case of (\ref{eq:CoordinateTransformation}) is given by:
\begin{equation}
\textbf{A} =
\left[\begin{array}{ccc}
\frac{\partial \rho'}{\partial \rho}           & \frac{1}{\rho}\frac{\partial\rho'}{\partial\varphi}           & \frac{\partial\rho'}{\partial z} \\
\rho'\frac{\partial\varphi'}{\partial \rho}    & \frac{\rho'}{\rho}\frac{\partial\varphi'}{\partial\varphi}    & \rho' \frac{\partial\varphi'}{\partial z} \\
\frac{\partial z'}{\partial \rho}              & \frac{1}{\rho}\frac{\partial z'}{\partial\varphi}             & \frac{\partial z'}{\partial z}
\end{array}\right]
=
\left[\begin{array}{ccc}
-n\left(\frac{r}{\rho}\right)^{n+1} & 0                  & 0 \\
0                                   & \frac{\rho'}{\rho} & 0 \\
0                                   & 0                  & -1
\end{array}\right].
\label{eq:JacobianMatrix}
\end{equation}
By inserting the above expressions into (\ref{eq:TransformedParameters}), the transformed relative material parameters for the core can be written in cylindrical coordinates as:
\begin{equation}
[\varepsilon^{\prime}_r]= [\mu^{\prime}_r] =
\left[\begin{array}{ccc}
n \varepsilon_{r \rho} & 0                                 & 0 \\
0                      & \frac{\varepsilon_{r \varphi}}{n} & 0 \\
0                      & 0                                 & \frac{\varepsilon_{r z}}{n} \left(\frac{r}{\rho'} \right)^{2+2/n}
\end{array}\right].
\label{eq:transformed_param}
\end{equation}
Furthermore, if we assume that the initial cylindrical shell has PML characteristics, that is, its relative constituent parameters are defined as $\varepsilon_{r\varphi} = \varepsilon_{rz}=1/\varepsilon_{r\rho}=a-jb$ (still assuming $[\varepsilon_r] = [\mu_r$]), we can write the transformed parameters as
\begin{equation}
[\varepsilon^{\prime}_r]= [\mu^{\prime}_r] =
\left[ \begin{array}{ccc}
\frac{n}{a-jb} & 0              & 0 \\
0              & \frac{a-jb}{n} & 0 \\
0              & 0              & \frac{a-jb}{n} \left(\frac{r}{\rho'} \right)^{2+2/n}
\end{array} \right].
\label{eq:CylindricalPML}
\end{equation}
It should be noted that we can derive the same material parameters if we, alternatively, demand $\varphi'=-\varphi$ instead of $z'=-z$ in the initial coordinate transformation. At this stage we are free to choose any real values for $a,b\in\mathbb{R}$ and any positive value for $n$; however, in this study we limit the analysis to the case $n=1$. Furthermore, we confine ourselves to the TM case with no harm to generality.

As we did in the case of an infinite planar slab, we consider three different routes approaching the ideal conjugate matched cylinder, namely the isotropic, nonactive and PML routes. Here, the isotropic route corresponds to isotropic shell parameters, that is, $\varepsilon_{r}=\mu_r= a-jb$. Clearly, according to (\ref{eq:transformed_param}) this corresponds to transformed parameters that are not isotropic. However, we will use the term ``isotropic'' also in this case to maintain harmony of the terminology with the planar slab example. In the nonactive scenario, as before for the planar case, the normal (radial) permittivity and permeability are assumed to be real $\left(\varepsilon_{r\rho'}'=\Re\left[\frac{1}{a-jb}\right]=\frac{a}{a^2+b^2}\right)$ while the other material parameters are given by (\ref{eq:CylindricalPML}). In the PML route, we again assume that the normal components of permittivity and permeability are slightly more lossy than it is dictated by the ideal PML rule (given in (\ref{eq:CylindricalPML})) in order to avoid possible numerical instabilities owed to the limiting nature of the ideal CML $\left(\varepsilon_{r\rho'}'=\frac{1}{a-jb}-j \delta \right)$.

\begin{figure}[ht]
\centering
\subfigure[]{\includegraphics[scale =0.17]{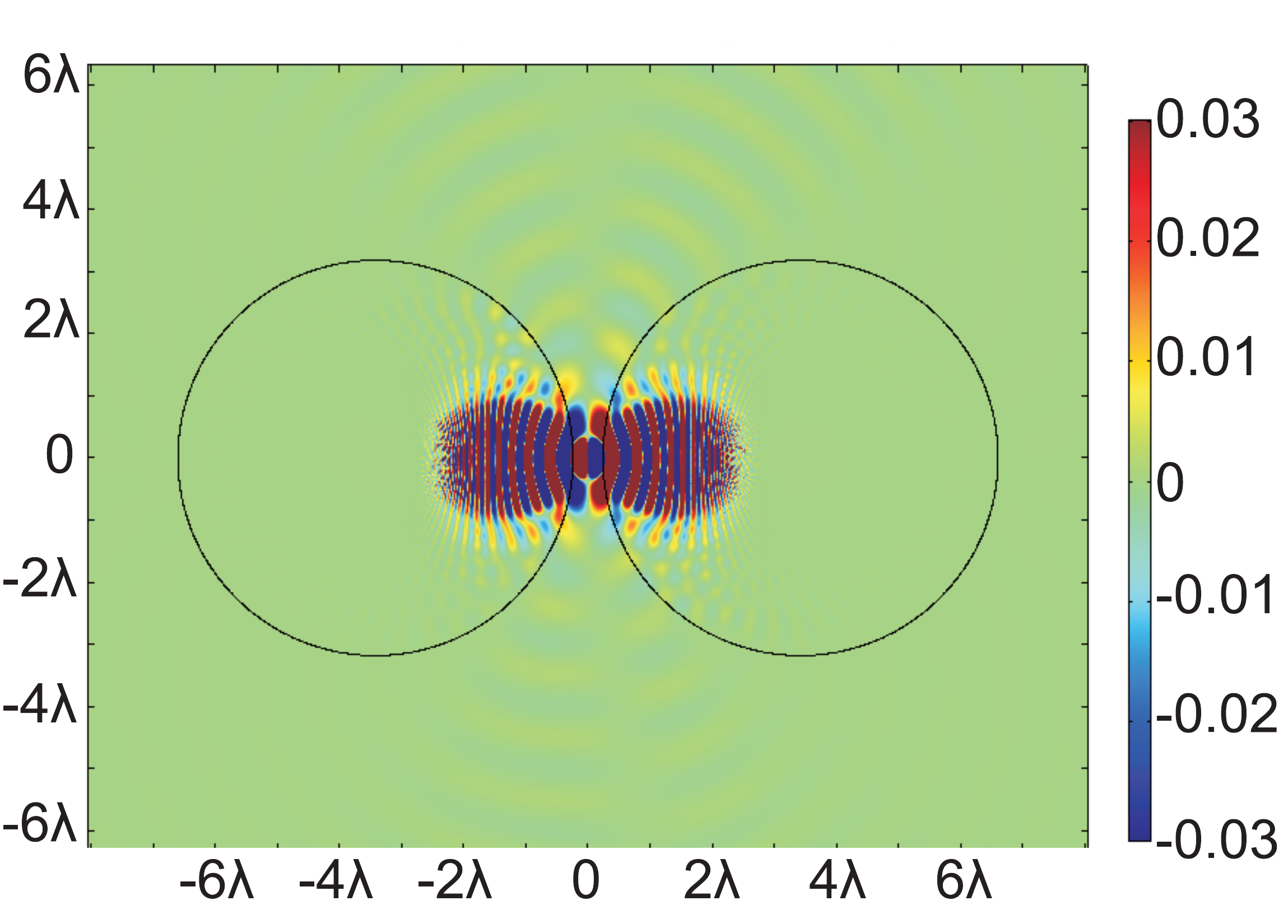}
   \label{fig:Fig11a}}
\subfigure[]{\includegraphics[scale =0.17]{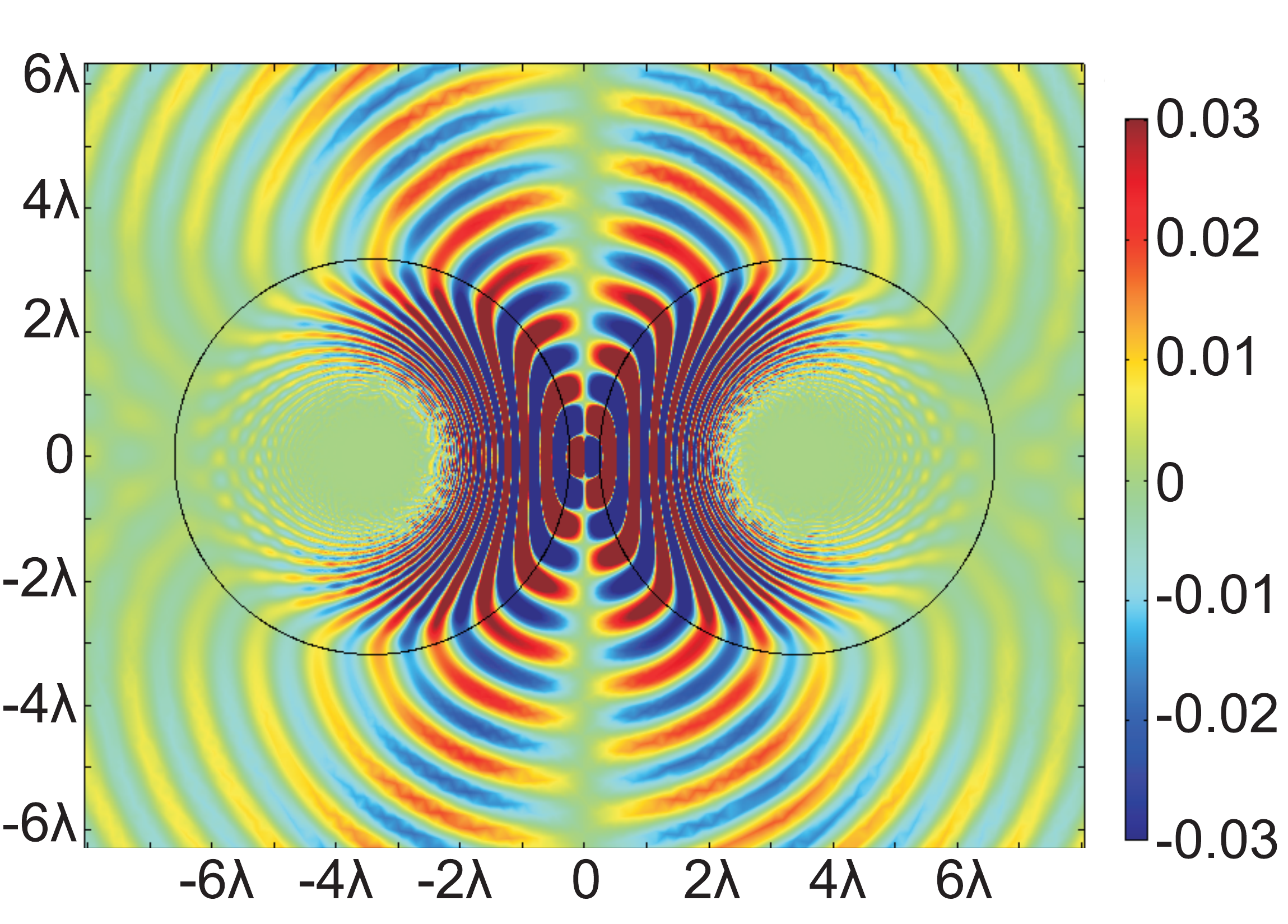}
   \label{fig:Fig11b}}
\subfigure[]{\includegraphics[scale =0.2]{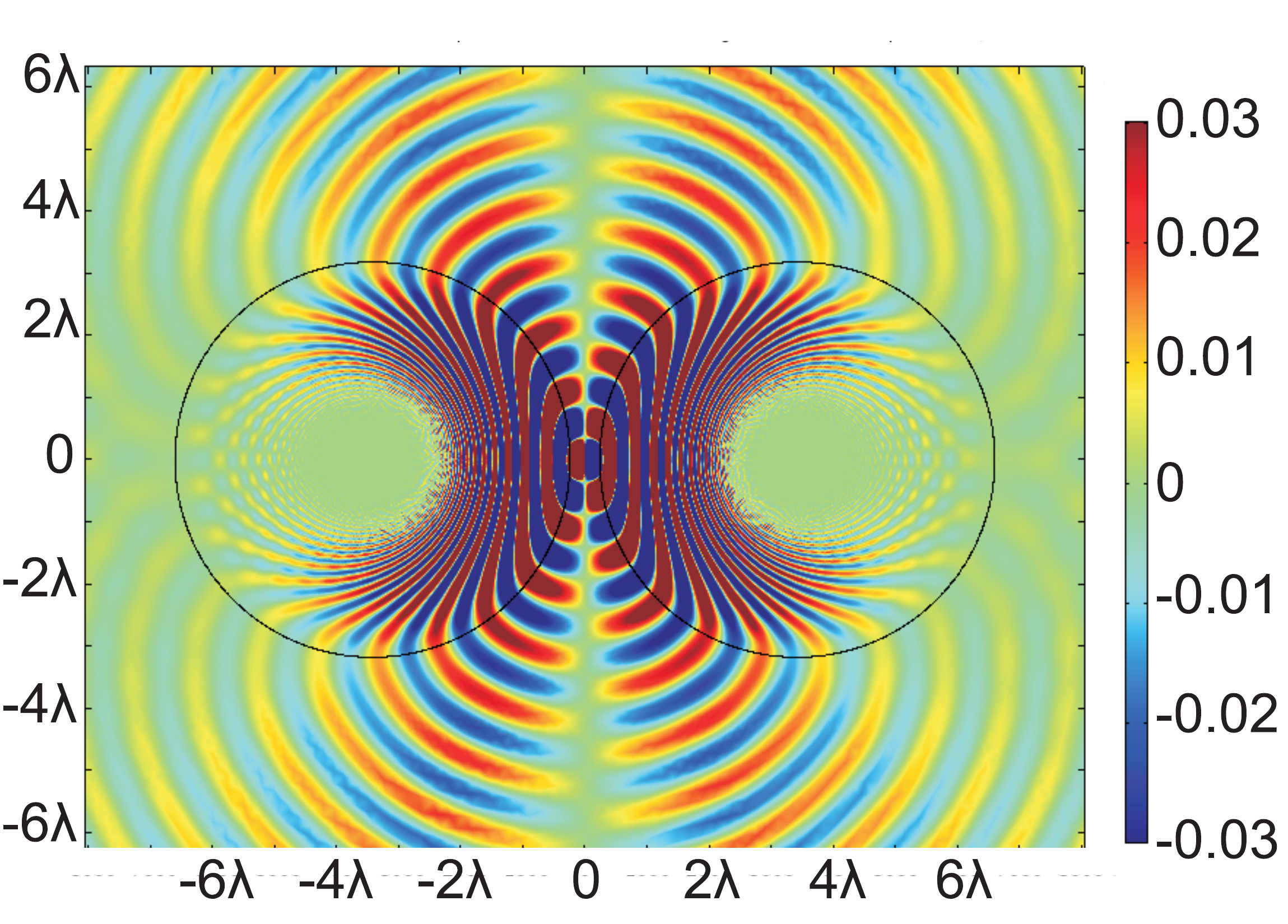}
   \label{fig:Fig11c}}
\caption{The spatial variation of the axial magnetic field $H_z(x,y)$ on the map $(x, y)$ for the three considered DPS routes: (a) the isotropic route $(\varepsilon^{\prime}_{r\rho}=a-jb)$, (b) the nonactive route $\left(\varepsilon^{\prime}_{r\rho}=\Re\left[1/(a-jb)\right]=a/(a^2+b^2)\right)$, and (c) the PML route $\left(\varepsilon^{\prime}_{r\rho}=1/(a-jb)-j\delta\right)$. Plot parameters: $a=2$, $k_0r=20$, $k_0d=3$, $\theta=90^\circ$, $b=0.05$, $\delta=0.005$, and $n=1$.}
\label{fig:Figs11DPS}
\end{figure}

\begin{figure}[ht]
\centering
\subfigure[]{\includegraphics[scale =0.17]{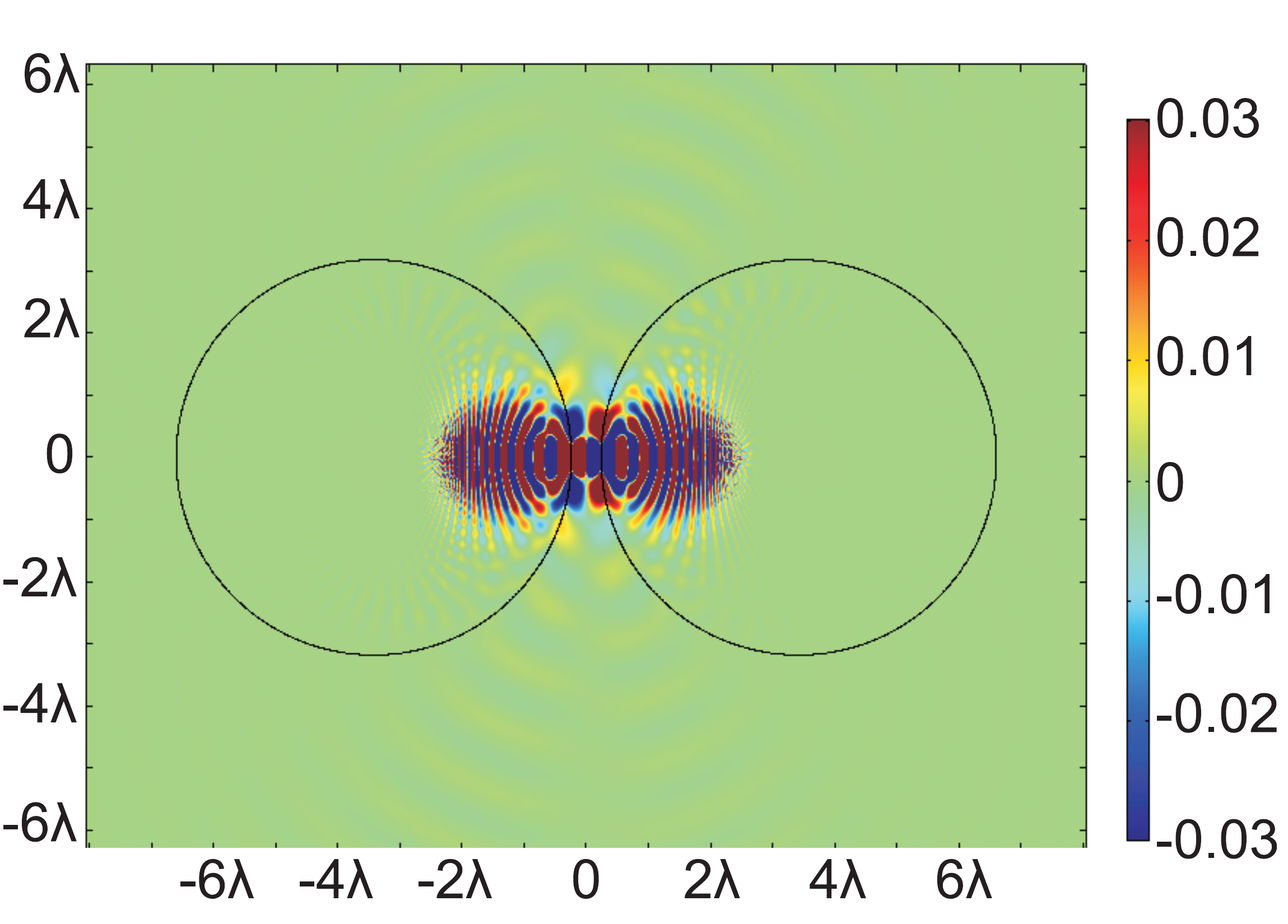}
   \label{fig:Fig11d}}
\subfigure[]{\includegraphics[scale =0.17]{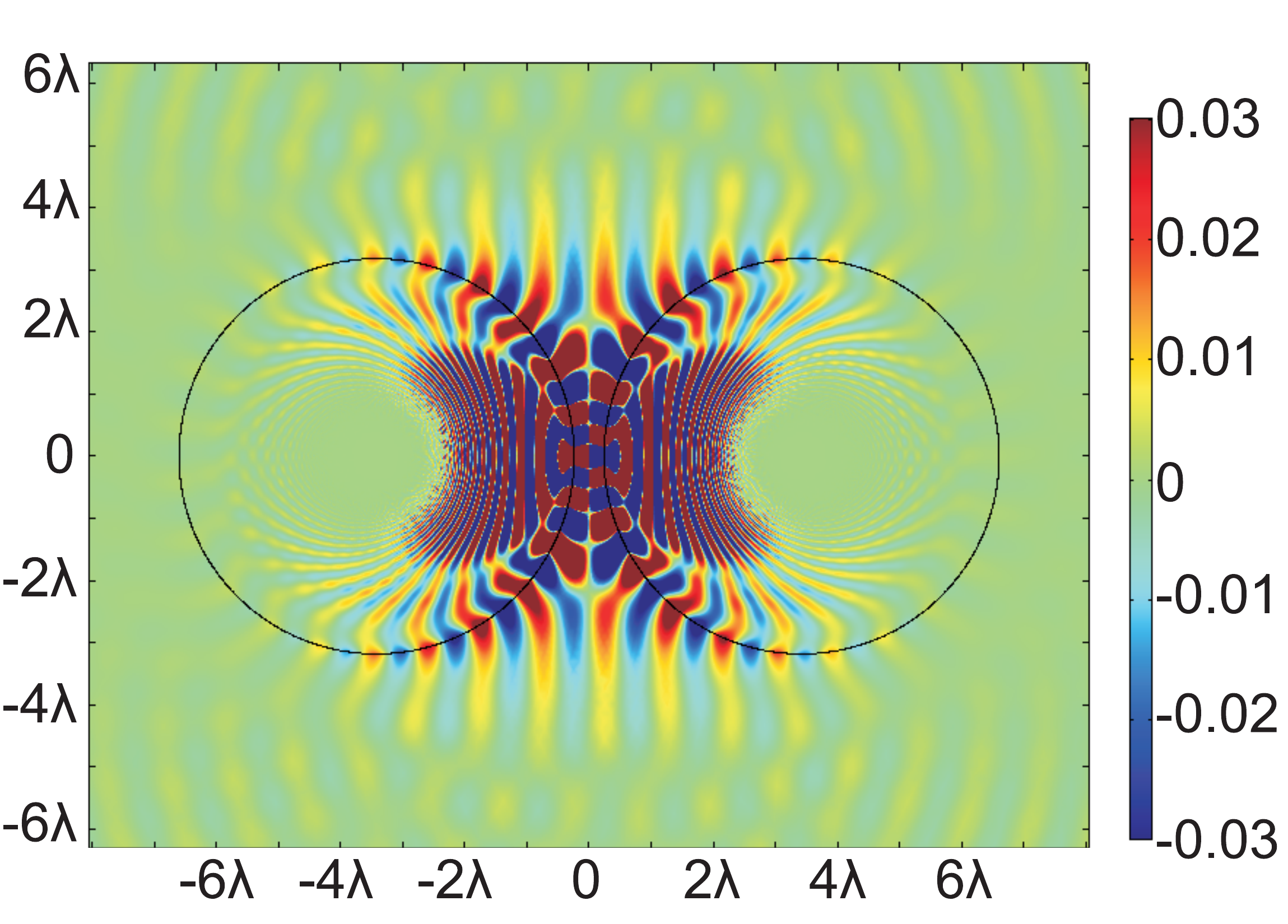}
   \label{fig:Fig11e}}
\subfigure[]{\includegraphics[scale =0.2]{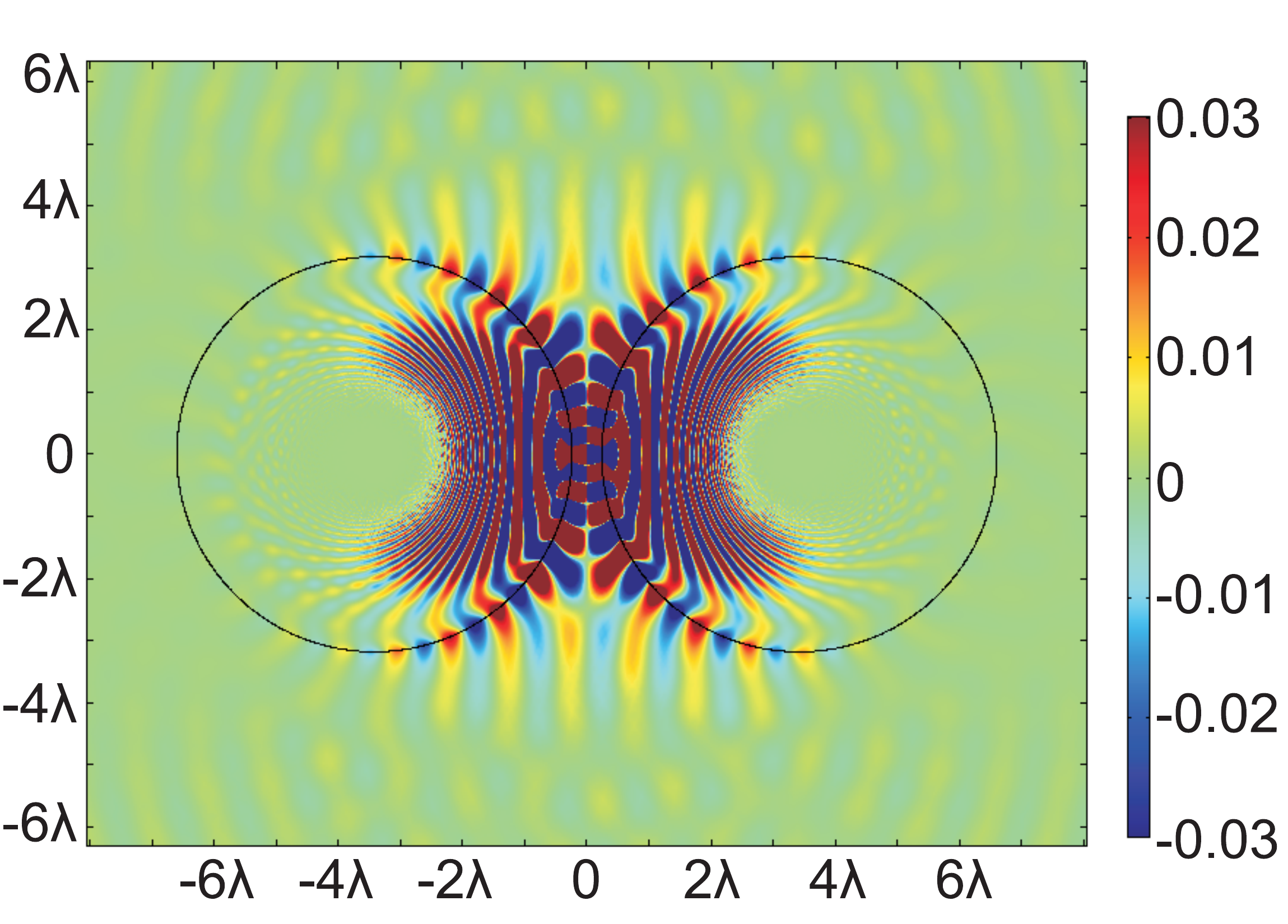}
   \label{fig:Fig11f}}
\caption{The spatial variation of the axial magnetic field $H_z(x,y)$ on the map $(x, y)$ for the three considered DNG routes: (a) the isotropic route $(\varepsilon^{\prime}_{r\rho}=a-jb)$, (b) the nonactive route $\left(\varepsilon^{\prime}_{r\rho}=\Re\left[1/(a-jb)\right]=a/(a^2+b^2)\right)$, and (c) the PML route $\left(\varepsilon^{\prime}_{r\rho}=1/(a-jb)-j\delta\right)$. Plot parameters: $a=-2$, $k_0r=20$, $k_0d=3$, $\theta=90^\circ$, $b=0.05$, $\delta=0.005$, and $n=1$.}
\label{fig:Figs11DNG}
\end{figure}

\subsection{Performance of the Proposed Structure}

In Figs.~\ref{fig:Figs11DPS} and \ref{fig:Figs11DNG}, we consider the case of two cylinders having material parameters corresponding to the three routes described in the previous subsection (``isotropic'', nonactive, and PML routes) and depicted in Fig.~\ref{fig:Fig8b}. We position the source line symmetrically in-between two cylinders of radius $r$ which are separated by distance $d$. When the source is that depicted in Fig.~\ref{fig:Fig4a}, namely a TM dipole line source, we again choose $\theta=90^\circ$ for maximal excitation of the system. In Figs.~\ref{fig:Fig11a}--\ref{fig:Fig11c} we consider the DPS (PML) cases with $a = 2$, while in Figs.~\ref{fig:Fig11d}--\ref{fig:Fig11f} the DNG (CML) ones with $a=-2$. By inspection of the contours, one directly infers that the DNG approach is much more efficient in terms of absorption: in the DNG case, practically all the power created by the source is absorbed by the cylinders. On the contrary, in all the DPS cases a considerable amount of power propagates away from the cylinders. Similar effect was observed for the DNG cylinders even if the dipole line excitation was replaced by a $z$-directed current line, despite the fact that the incident field is nonzero along the $\hat{\textbf{y}}$ axis. The difference between Figs.~\ref{fig:Fig11b}, \ref{fig:Fig11c} and between Figs.~\ref{fig:Fig11e},\ref{fig:Fig11f} is negligible due to small losses $b$ combined with the choice of fairly large $|a|$. Figs.~\ref{fig:Fig11e},\ref{fig:Fig11f} show that surface plasmon modes are strongly excited due to smaller overall losses compared to the other cases. In the nonactive and PML routes with $a=-2$ shown in Figs.~\ref{fig:Fig11e} and \ref{fig:Fig11f}, surface plasmon modes are strongly excited on the surface of the cylinders which does not happen for the isotropic DNG case shown in Fig.~\ref{fig:Fig11d} due to higher losses in the radial material parameters compared to the other cases.

\begin{figure}[ht]
\centering
\subfigure[]{\includegraphics[scale =0.5]{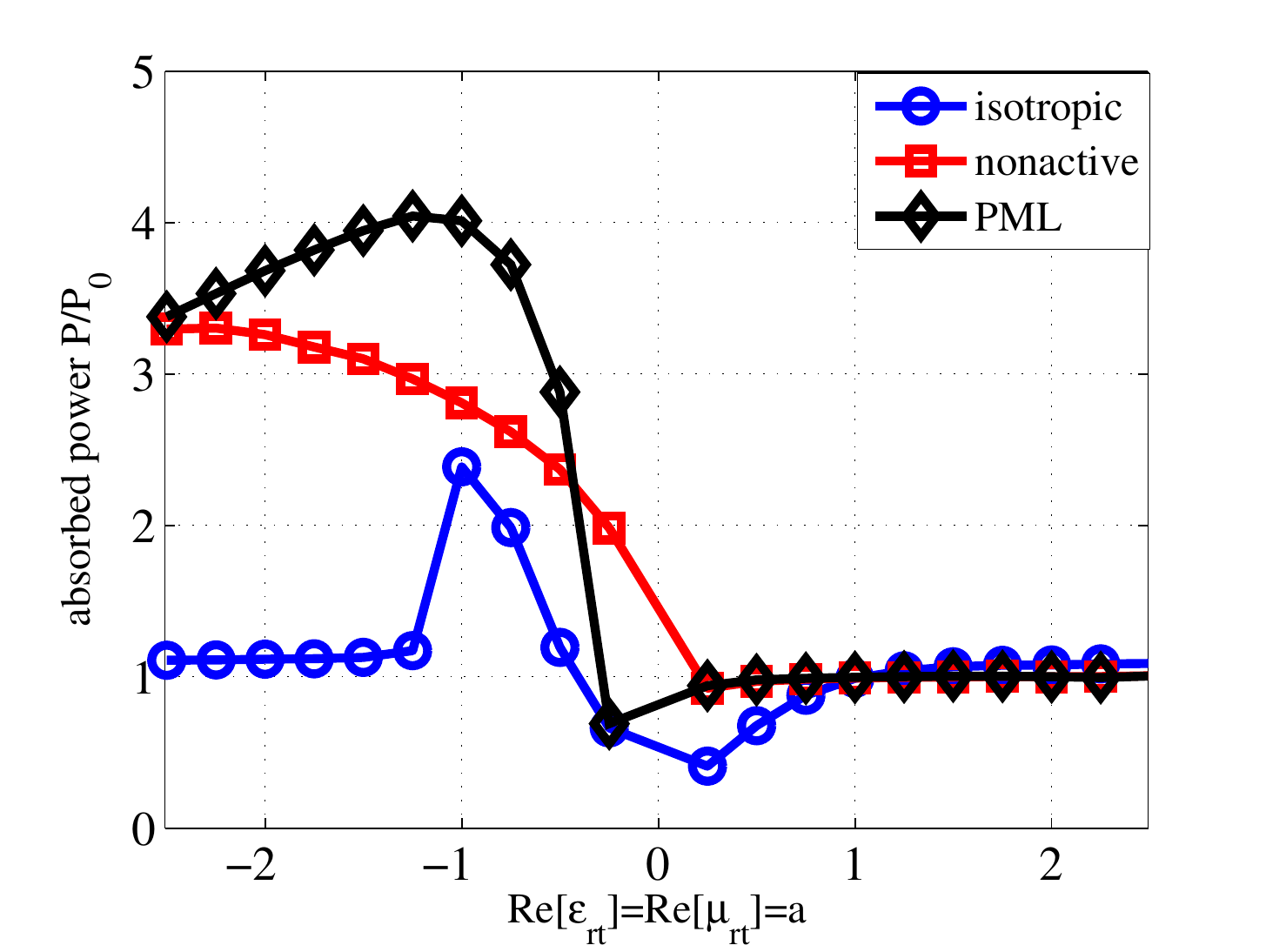}
   \label{fig:Fig12a}}
\subfigure[]{\includegraphics[scale =0.5]{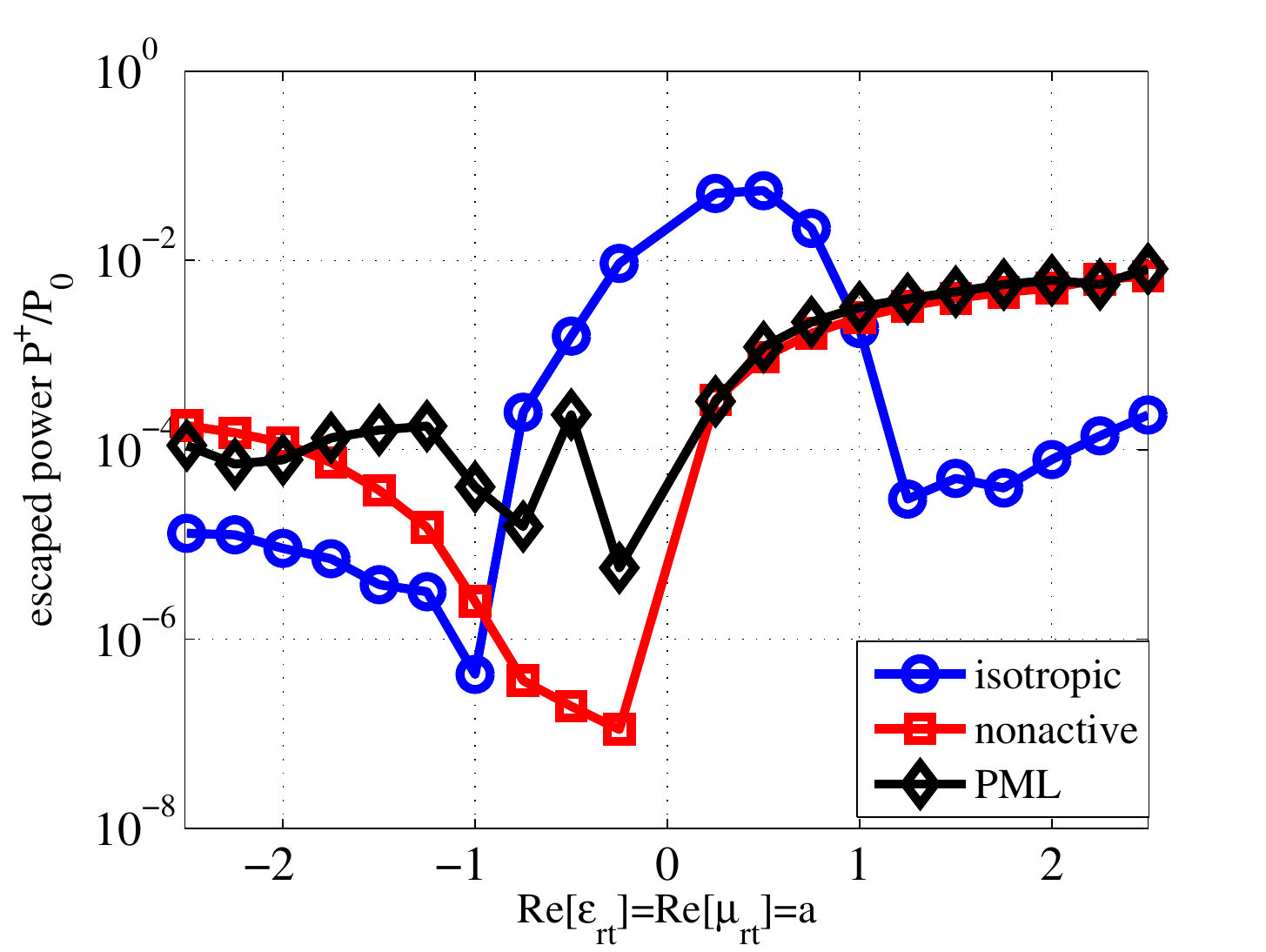}
   \label{fig:Fig12b}}
\caption{(a) The normalized absorbed power $P/P_0$ by the two cylinders and (b) the escaping power $P^+/P_0$ from the system as functions of the material parameter $a$ in the three different scenarios (isotropic, nonactive, PML). Plot parameters: $k_0r=20$, $k_0d=3$, $\theta=90^\circ$, $b=0.05$, $\delta=0.005$.}
\label{fig:Figs12}
\end{figure}

In Fig.~\ref{fig:Fig12a} we show the variations of the absorbed power $P$ normalized by $P_0$, which corresponds to the absorbed power in the DPS-PML case with $a=1$, with respect to $a$. Given the fact that the conventional PML absorbs only propagating waves, the physical meaning of $P_0$ is identical to the corresponding one of the planar slab cases. Regardless of the considered route (isotropic, nonactive, {PML}), there is a substantial switch in absorption efficiency from $a<0$ (DNG case) to $a>0$ (DPS case): the absorbed power $P$ is always much higher for $a<0$. For positive values of $a$, all routes lead to similar results. The higher absorption in the isotropic shell case is simply due to the lossy radial component (as opposed to nonactive/active $\varepsilon_{r\rho}$ used in the other two cases). When it comes to the isotropic shell case, the maximal absorption is achieved when $a=-1$, since the power drops quickly as $a$ deviates from $-1$ in either direction; similar behavior has been observed also for the planar slab case, as well as for the double-negative sphere \cite{SphericalPaper}. It should be noted that at $a=\pm 1$ practically the only difference between the material parameters given by the three scenarios are the losses in the radial permittivity and permeability. Furthermore, it should be stressed that in the nonactive DNG case, the maximum absorbed power (exhibited at $a\cong -2.5$) is smaller than in the PML scenario but larger than in the isotropic shell case, as could be expected. Similarly to the planar case, the highest $P$ is recorded for the PML route. In Fig.~\ref{fig:Fig12b} we present the escaped power $P^+$ with respect to the real part of the transversal relative permittivity $a$. $P^+$ denotes the power carried away by the surviving fields at $\rho^{\prime}\rightarrow +\infty$. The values of $P^+$ for $a<0$ are greatly diminished compared to the DPS cases, a feature that demonstrates the higher absorption effectiveness of the DNG structures. For $|a|>1$, the escaped power is the smallest for the isotropic scenario which is, again, due to  higher losses compared to the other cases.

\begin{figure}[ht]
\centering
\subfigure[]{\includegraphics[scale =0.28]{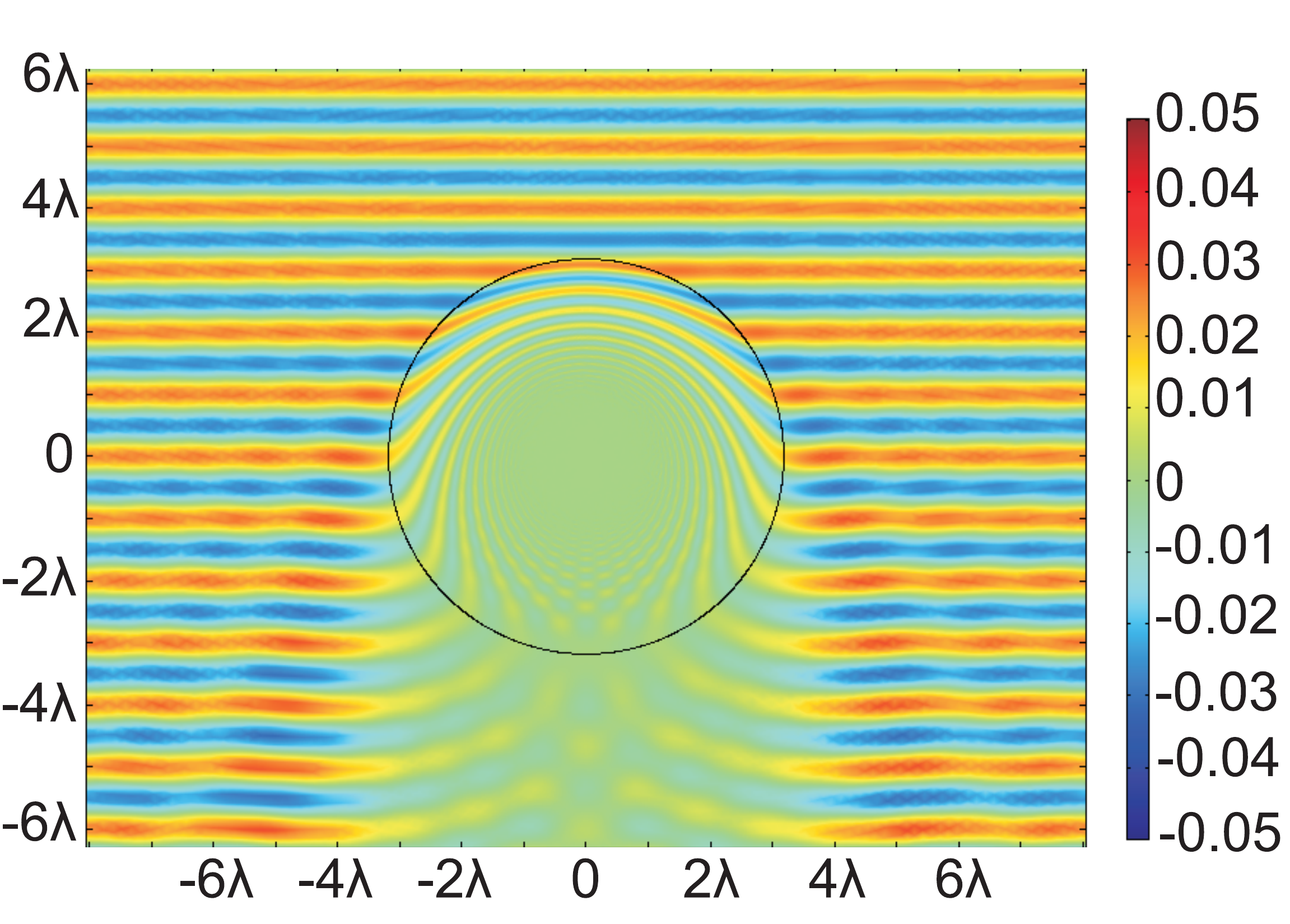}
   \label{fig:Fig13a}}
\subfigure[]{\includegraphics[scale =0.28]{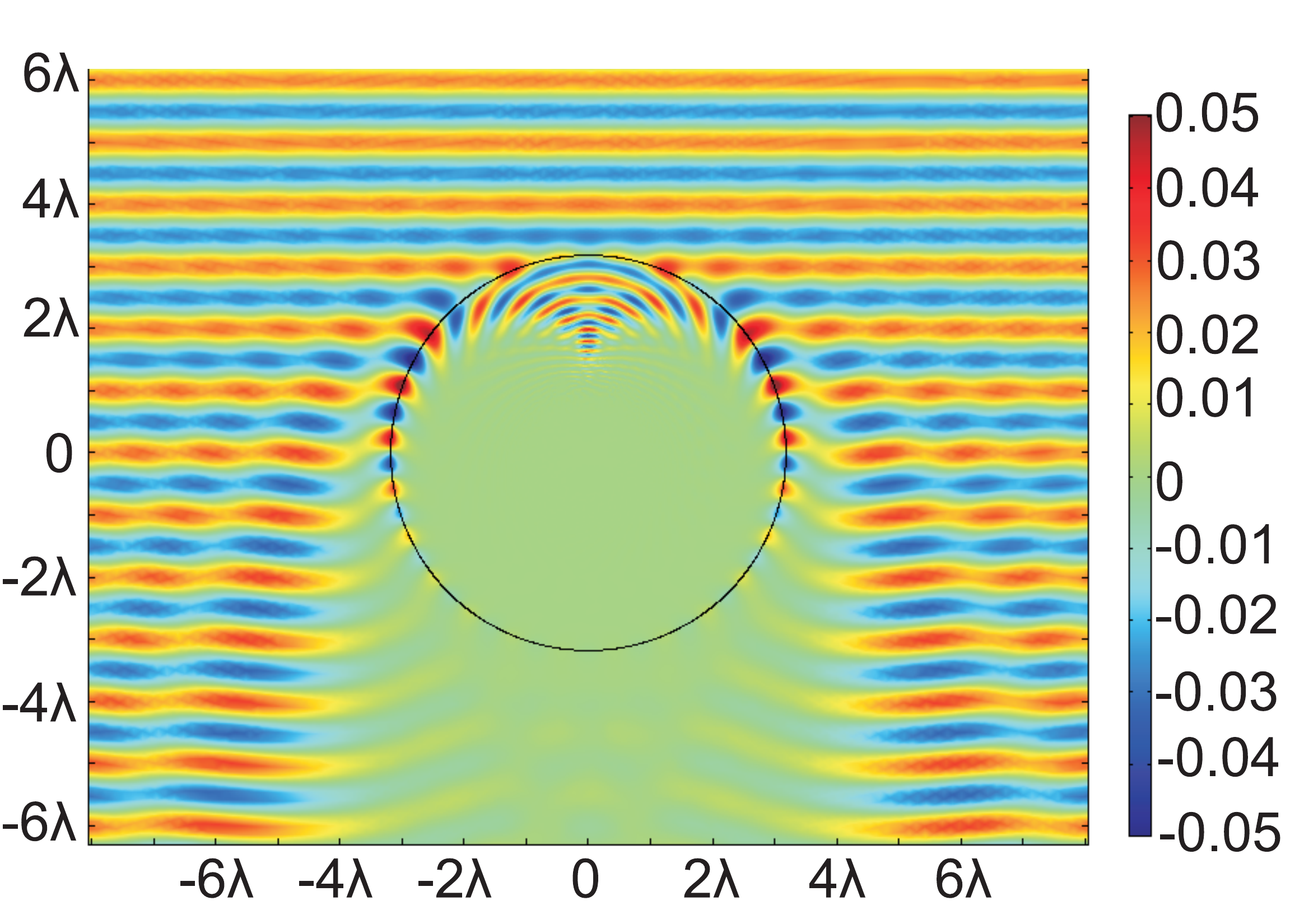}
   \label{fig:Fig13b}}
\caption{Plane-wave illumination. The spatial variation of the axial magnetic field $H_z(x,y)$ on the map $(x, y)$ for the PML route $\left(\varepsilon^{\prime}_{r\rho}=1/(a-jb)-j\delta\right)$ with (a) $a=2$ and (b) $a=-2$. Plot parameters: $k_0r=20$, $k_0d=3$, $b=0.05$, $\delta=0.005$, and $n=1$.}
\label{fig:Figs13}
\end{figure}

Finally, in Figs.~\ref{fig:Figs13} we examine absorption effectiveness of double-negative conjugate matched cylinders illuminated by a single propagating  plane wave. According to the results of Ref.~\onlinecite{SphericalPaper}, in the limiting case of the ideal CML, the effective absorption width can be arbitrarily large, although the sources of illumination are infinitely far from the cylinder. We consider only the PML route for $a=2$ (DPS, Fig.~\ref{fig:Fig13a}) and $a=-2$ (DNG, Fig.~\ref{fig:Fig13b}). Note that in both cases, there are practically no reflections from the cylinder and a huge shadow is formed behind. Furthermore, resonant plasmonic modes are excited on the surface of the cylinder in the DNG case, which increases the shadow due to stronger field coupling. Although the excitation is a single propagating plane wave, coupling to predominantly evanescent resonant modes of the absorbing cylinder is possible because the surface is curved. In this particular example, the power absorbed by the cylinder is about $18\%$ higher in the DNG case compared to the DPS case (the latter case corresponds to the ideal black body as defined in Ref.\cite{GK}). In other words, even in the absence of incident evanescent plane-wave modes we can achieve improved absorption using DNG cylinders as compared to the perfectly non-reflecting DPS cylinder, thanks to the aforementioned activation of resonant surface plasmon modes. However, for the same material parameters, the effect is quite modest compared to the increased power absorption when evanescent wave modes are strongly present in the spatial spectrum of the source.

\section{Practical Realization of the CML Concept}
\label{sec:Realization}

As is clear from the above results, two scenarios are promising for realizations of the proposed conjugate matched bodies: the nonactive route and the PML route. In the nonactive route, the required material parameters correspond to well-studied uniaxial double-negative (backward-wave) metamaterials \cite{MetaReview}. A wide variety of micro- and nanostructures which realize double-negative response in different frequency ranges have been proposed and studied in the literature. For the new application proposed here, the materials should have a controllable degree of anisotropy: in particular, the case when the absolute values of the tangential components are much larger than those of the normal components is of interest. With this in view, one of the most promising  topologies is the so-call fishnet structure \cite{FishnetMetamaterials, MetaReview, SlabWire, PolarizationIndependent}. This is a multilayer structure where thin metal sheets with periodically positioned holes (usually of the square or round shape) are separated by dielectric spacers.

While the response in the tangential plane is strongly modified by large induced currents flowing along the metal sheets, the response to fields along the normal direction is modified only due  to quasi-static electric polarization of thin sheets. In the microwave domain, the normal components of the material parameters usually have positive real parts. To achieve the necessary negative permittivity value of the normal component, it is possible to insert an array of thin metal wires passing through the holes in the fishnet layers. If the wires are made of a good conductor, the microwave response will correspond to a negative and nearly real value of the normal component of the permittivity tensor, exactly as required for the CML realization.

At infrared and optical frequencies, it is possible to exploit the negative permittivity of metals from which the fishnet layers can be fabricated. In this case, metal wires are not needed, and the values of the normal component of the permittivity can be controlled by choosing the thickness ratio of the metal and dielectric layers in the fishnet structure.
In particular, if rectangular holes are etched in a layer of Magnesium fluoride (MgF$_2$) sandwiched between silver slabs, one can achieve negative refractive index \cite{SilverMagnesiumFluoride} at the optical frequencies under the necessary homogenization conditions \cite{OurFishnet}. Furthermore, nonlinearity of liquid crystals can  be  exploited in order to develop tunable fishnets metamaterials whose reflection/transmission (and accordingly their effective $\varepsilon_{rt}, \mu_{rt}$) is externally controlled. More generally speaking, it is well known that if we insert metallic particles such as cylindrical pins, strips, spirals or flakes, which are having negative permittivity in the visible spectrum \cite{MetallicNegativePermittivity}, we can  \cite{EffectiveFormula} obtain materials with $\Re[\varepsilon_{rn}]<0$.

The conclusion that the performance of nonactive-route CML improves when the absolute value of the real part of the tangential permittivity increases (Section~\ref{sec:Planar}) is very important for choosing practically realizable material parameters. As was discussed above, high absorption of energy from evanescent fields requires that the overall losses in the medium are small (this is in fact clear already from eq.~(\ref{conj_match})). Realization of passive low-loss DNG metamaterials is a challenge, although some successful approaches are known, e.g. \cite{PMMLossesA}. However, for the nonactive PML with large magnitudes of negative $a$, high absorption is observed even when the imaginary part of the tangential permittivity $b$ is not very small and corresponds to practically realizable structures. The normal component of the permittivity in this case corresponds to epsilon-near-zero materials, which are also known to be practically  realizable \cite{Liisi,Engheta_ENZ}.

The required overall small level of losses can achieved also using active (pumped) structures. This can be done  by using active parts such as optically pumped laser dyes to balance the losses of the background structure \cite{PMMLossesB}. In this way, one can realize an effective DNG medium possessing additional negative normal components ($\Re[\varepsilon_{rn}]<0$) with controllable loss factors of the tangential components: $\Im[\varepsilon_{rt}], \Im[\mu_{rt}]<0$. Active control of the normal permittivity and permeability components  can be used to implement also the active PML route towards CML. Based on the fishnet-wire medium approach described above, instead of metallic rods or spirals one can insert nano-generators \cite{ActivePMMA} or pumps \cite{ActivePMMB} into the holes of the background structure which can provide energy to the system and make the normal component of the effective parameters active.

\section{Conclusions}

In this paper, we have introduced the concept of  double-negative \emph{conjugate matched layer} (CML), which has the property of acting as the ideal sink for energy of electromagnetic fields. Arbitrary propagating waves illuminating such layer produce no reflections and are fully absorbed in the medium. Evanescent fields, existing in the vicinity of small sources or scatterers, induce very strong resonant fields, oscillating in the vicinity of the CML surface. The energy stored in the reactive evanescent modes sinks into the layer and is transformed into heat or delivered to loads in its constitutive elements.

There is an analogy of this phenomenon and the circuit-theory concept of the ideal voltage or current source. An ideal voltage source delivers infinite power to the load in the limit of infinitely small resistive load [eq.~(\ref{conj_match})], because the current through a resistor tends to infinity when the resistance tends to zero. Likewise, for every evanescent mode the sum of the wave impedances of fields in free space and that in the CML is real and can be made very small. Thus, the layer dramatically enhances the near fields created by given sources and sacks their energy into the material layer. This phenomenon can be compared also with the Purcell  effect of enhancement of spontaneous emission rate of small sources due to the presence of resonant and efficiently radiating bodies in the vicinity of the source \cite{Nov,Alu_Eng}. However, resonant cavities or antennas used to enhance radiation from a small emitter work only at one or a few resonant modes of the resonator. The introduced CML concept brings the Purcell effect to the limit of extracting  \emph{all} available energy from a small source. In this paradigm, every mode of the source field resonates with a corresponding mode of the CML, so that the complete system of the source and the absorbing body is tuned for the optimal power extraction from \emph{all} modes of the fields created by the source.

Although the ideal double-negative conjugate matched layer requires active inclusions for its practical realization \cite{PML_active} (the normal components of the material parameters have the imaginary parts corresponding to active media), the flexibility of the CML concept suggests a more practical possibility for realizations, where only passive materials are used: the nonactive route, described in this paper, requires double-negative \emph{lossy} values of the tangential components and low-loss negative values of the normal components of the permittivity and permeability matrices. The results of extensive studies   of various possibilities to design and realize double-negative (backward-wave) materials can be used to realize the proposed structures. While the earlier introduced isotropic double-negative design \cite{SphericalPaper} requires very small levels of losses, which are probably not possible to achieve in practice, at least with the use of only passive materials, the radially uniaxial double-negative structure proposed here appears to offer possibilities to overcome this limitation. We hope that  the results of this study will have interesting and important implications for a wide variety of applications: in light harvesting, thermal emission control, heating and cooling devices, stealth, decoupling of radiators, receiving and transmitting antennas, and other microwave and optical devices.

\medskip

\begin{acknowledgments}
S.\ I.\ Maslovski acknowledges financial support under ``Investigador FCT (2012)'' grant.
\end{acknowledgments}

\appendix*
\label{AnnexA}
\section{Expression for the Incident Field Created by the Dipole Line}

In the spectral domain, the free-space vector potential created by the line of dipoles shown in Fig.~\ref{fig:Fig4a} can be written for this 2D problem as
\begin{equation}
\textbf{A}_{\rm 0,inc}(k_x,k_y)=\frac{jk_0\eta_0\textbf{p}_le^{-jk_xg}}{k_x^2+k_y^2-k_0^2},
\label{eq:MagneticVectorPontential}
\end{equation}
where $\textbf{p}_l$ is the electric dipole moment linear density. Taking the inverse Fourier transform $\left(\mathbfcal{A}_{\rm 0,inc}(x,k_y)=\frac{1}{2\pi}\int_{-\infty}^{+\infty}\textbf{A}_{\rm 0,inc}(k_x,k_y)e^{-jk_xx}dk_x\right)$ with respect to $k_x$ we obtain
\begin{equation}
\mathbfcal{A}_{\rm 0,inc}(x,k_y)=\frac{jk_0\eta_0\textbf{p}_l}{2}\frac{e^{-j\beta_0|x+g|}}{\beta_0},
\label{eq:TransformedMagneticVectorPontential}
\end{equation}
where $\beta_0=\beta_0(k_y)=-j\sqrt{k_y^2-k_0^2}$.  When $k_y=0$, (\ref{eq:TransformedMagneticVectorPontential}) reduces to the well-known expression for the vector potential of a uniform electric dipole moment sheet. The  magnetic field of the same line of dipoles is found using:
\begin{equation}
\mathbfcal{H}_{\rm 0,inc}(x,k_y)=\frac{1}{\mu_0}\nabla\times\mathbfcal{A}_{\rm 0,inc}(x,k_y),
\label{eq:Potential2Field}
\end{equation}
where $\nabla=\hat{\textbf{x}}\frac{\partial}{\partial x}-\hat{\textbf{y}}jk_y$ (recall that there is no dependence on $z$).  The result reads
\begin{equation}
\mathbfcal{H}_{\rm 0,inc}(x,k_y)=\hat{\textbf{z}}\frac{j\omega p_l}{2}e^{-j\beta_0|x+g|}\left[\frac{k_y}{\beta_0}\cos\theta-\sin\theta\right],
\label{eq:TransformedMagneticField}
\end{equation}
where $p_l = |\_p_l|$. Because the vector $\_p_l$ lies in the $xy$ plane, the incident magnetic field has just a single component along the $z$ axis, i.e., the incident waves are TM-polarized plane waves.

\end{document}